\documentclass{article}

\usepackage{PRIMEarxiv}

\usepackage[utf8]{inputenc} 
\usepackage[T1]{fontenc}    
\usepackage{hyperref}       
\usepackage{url}            
\usepackage{booktabs}       
\usepackage{amsfonts}       
\usepackage{nicefrac}       
\usepackage{microtype}      
\usepackage{lipsum}
\usepackage{fancyhdr}       
\usepackage{graphicx}       

\usepackage{algorithmic, algorithm2e, color}
\usepackage{exscale,relsize}
\usepackage{amsfonts}
\usepackage{mdwlist}

\usepackage{subcaption}

\usepackage{tikz}
\usepackage{multirow}
\usepackage{array} 
\usepackage{url}
\usepackage{listings}
\usepackage{dsfont}
\usepackage{enumitem}

\usepackage{xcolor}
\usepackage{paralist}

\newcommand{\system}{\texttt{Thermanator}\xspace}
\newcommand{\combine}{\texttt{AcuTherm}\xspace}
\newcommand{\degC}{^{\circ}C}
\newcommand{\degK}{K} 
\newcommand{\acutherm}{\texttt{AcuTherm}\xspace}
\newcolumntype{P}[1]{>{\centering\arraybackslash}p{#1}}

\newtheorem{assumption}{Assumption}

\graphicspath{{fig/}}

\makeatletter
\newcommand\notsotiny{\@setfontsize\notsotiny\@vipt\@viipt}
\makeatother

\newcommand{\surf}{\textsf{Shoulder-Surfing}\xspace}
\newcommand{\lunch}{\textsf{Lunch-Time}\xspace}
\newcommand{\acoust}{\textsf{Acoustic Emanations}\xspace}
\newcommand{\vibrat}{\textsf{Keyboard Vibrations}\xspace}
\newcommand{\hybrid}{\textsf{AcuTherm}\xspace}
\newcommand{\attack}{\textsf{Thermanator}\xspace}

\newcommand{\surfa}{\textsf{Shoulder-Surfing Attacks}\xspace}
\newcommand{\luncha}{\textsf{Lunch-Time Attacks}\xspace}
\newcommand{\attacka}{\textsf{Thermanator Attacks}\xspace}
\newcommand{\acutherma}{\textsf{AcuTherm Attacks}\xspace}

\newcommand{\hybrida}{\textsf{Hybrid Attacks}\xspace}

\ifdefined\showchanges
\newcommand{\changed}[1]{\color{blue}{#1}\color{black}\xspace}
\else
\newcommand{\changed}[1]{#1\xspace}
\fi

\title{Thermal (and Hybrid Thermal/Audio) Side-Channel Attacks on Keyboard Input}

\author{
  Tyler Kaczmarek \\
  MIT Lincoln Labs \\
  MA, USA \\
  \texttt{Kaczmarek@ll.mit.edu} \\
   \And
  Ercan Ozturk \\
  University of California, Irvine \\
  CA, USA \\
  \texttt{ercano@uci.edu} \\
  \And
  Pier Paolo Tricomi \\
  University of Padua \\
  Padua, Italy \\
  \texttt{tricomi@math.unipd.it} \\
  \And
  Gene Tsudik \\
  University of California, Irvine \\
  CA, USA \\
  \texttt{gene.tsudik@uci.edu} \\
}

\begin{document}
\maketitle

\begin{abstract}
To date, there has been no systematic investigation of thermal profiles of keyboards, 
and thus no efforts have been made to secure them. This serves as our main motivation 
for constructing a means for password harvesting from keyboard thermal emanations. 
Specifically, we introduce \system: a new post-factum insider attack 
based on heat transfer caused by a user typing a password on a typical external (plastic) keyboard. 
We conduct and describe a user study that collected thermal residues from $30$ 
users entering $10$ unique passwords (both weak and strong) on $4$ popular commodity keyboards. 
Results show that entire sets of key-presses can be recovered by non-expert users as late as $30$ seconds
after initial password entry, while partial sets can be recovered as late as $1$ minute after entry. 
However, the thermal residue side-channel lacks information about password length, duplicate key-presses, and key-press ordering. To overcome these limitations, we leverage keyboard acoustic emanations and combine the two to yield \combine, the first hybrid side-channel attack on keyboards. 
\combine significantly reduces password search without the need for any training on the victim's typing. We report results gathered for many representative passwords based on a user study involving 19 subjects.

The takeaway of this work is three-fold: (1) using plastic  keyboards 
to enter secrets (such as passwords and PINs) is even less secure than previously recognized, (2) post-factum
thermal imaging attacks are realistic, and (3) hybrid (multiple side-channel) attacks are both realistic and effective. 
\end{abstract}


\section{Introduction}
\label{sec:intro}
Insider attacks are very common, estimated to account for ${\approx}28\%$ of all 
electronic crimes in industry~\cite{mickelberg2014us}. This includes some high-profile attacks, 
such as the 2014 Sony hack~\cite{robb2014sony}. The danger of insider attacks mainly stems 
from  the fact that insiders often have privileged access. More importantly, 
insider attackers might be able to surreptitiously obtain
credentials of their coworkers/colleagues, thus allowing {\em lateral movement}. 
Such credential theft attacks occur because the
attacker's current privileges are insufficient to complete planned malicious 
tasks~\cite{nurse2014understanding}, or the attacker's goal is to  
evade accusations by putting the blame on others.

Since passwords are still the most common type of credentials, they are a major 
target for insider attackers. The danger of 
password compromise attacks are amplified because: \textit{(1)} people often use 
the same password on multiple systems, and \textit{(2)} most systems support ``Forgot password?'' 
schemes (to update or recover passwords) using 
the original email account, which is often in a logged-in state due to convenience.

At the same time, it is well known that the security 
of a system is based on its weakest link. Furthermore, it is often assumed that the involvement of a fallible 
(or simply gullible) human user corresponds to this weakest link, e.g., as in \surf and \lunch attacks~\cite{conti2020auth}. 
However, other insider attacks that focus on stealing passwords by compromising the user environment, 
e.g., \acoust~\cite{asonov2004keyboard,zhuang2009keyboard,compagno2017don}
or \vibrat~\cite{owusu2012accessory}, show that the weakest link is a consequence 
of certain laws of Physics in the form of side-channels.

Although side-channels can be effective (with optimal environmental conditions, equipment and time), 
information gleaned from them are usually incomplete, thus still leaving the attacker with a sizeable password search 
space. One intuitive way to reduce the attacker's search space is to combine multiple side-channels.

In this paper, we introduce \attack, a novel thermal residue side-channel attack on passwords entered on 
external keyboards, and evaluate its efficacy. We then supplement the thermal side-channel with its audio
counterpart (via keyboard acoustic emanations) to yield \acutherm, the first hybrid side-channel attack. 
Sections~\ref{subsec:heat} and~\ref{subsec:acoustics}, overview these two side-channels.

\subsection{Heat Transfer \& Thermal Emanations}\label{subsec:heat}
Any time two objects with unequal temperatures come in contact with each other, an 
exchange of heat occurs. This is unavoidable. Being warm-blooded, human beings naturally prefer
environments that are colder than their internal temperature. Because of this heat disparity, 
it is inevitable that we leave thermal residue on numerous objects that we routinely touch, especially, with bare fingers. 
Furthermore, it takes time for these heated objects to cool off and lose heat energy imparted by 
human contact. It is both not surprising and worrisome that this includes our interactions with 
keyboards that are used for entering sensitive private information, such as passwords. 

Based on this observation, we consider a mostly unexplored attack space where heat 
transfer and subsequent thermal residue can be exploited by a clever adversary to steal 
passwords from a keyboard some time after it was used for password entry. The main
distinctive benefit of this attack type is that adversary's real time presence is not required.
Instead, a successful attack can occur with after-the-fact adversarial presence:  
as our results show, many seconds later.

While there has been some prior work on using thermal emanations to crack PINs, mobile 
phone screen-locks and  opening combinations of vaults/safes~\cite{SafeCracking,andriotis2013pilot, 
abdelrahman2017stay,mowery2011heat}, this work represents the first comprehensive investigation of 
human-based thermal residues and emanations of external computer keyboards.

\subsection{Keyboard Acoustics}\label{subsec:acoustics}
Acoustic side-channel attacks rely on unique sounds produced during the 
processing of a secret to gather information. Previous work includes recovering 
various types of secrets, such as printed texts~\cite{backes2010acoustic}, 
3D-printed objects~\cite{faruque2016acoustic} and cryptographic keys~\cite{genkin2014rsa}. 

Generally, acoustic side-channel attacks against password entry are based on  the sounds 
produced by pressed keys on a keyboard. These sounds were shown to be distinct 
\cite{asonov2004keyboard}, allowing an attacker to differentiate among pressed keys and thus recover 
passwords, even in a remote VoIP setting ~\cite{compagno2017don}. In addition, 
inter-keystroke timings can be used to reduce password 
search space~\cite{song2001timing,foo2010timing} via various
statistical techniques to determine likely candidate key-pairs. If dictionary 
passwords are used, methods similar to those in 
\cite{zhuang2009keyboard} can be used due to the underlying base language 
properties. For random passwords, dictionaries are not applicable, since they 
lack the structure that can be used to reduce password 
search space.~\cite{halevi2015keyboard} investigates this phenomenon
and suggests a brute-force password search mechanism 
based on 5 best-guesses for each key in the password, similar to 
the one in~\cite{compagno2017don}.

Unfortunately, acoustic side-channels often involve a lengthy training phase
(i.e., profiling) of victim's typing style  and provide incomplete information on 
the target secret, e.g., inter-keystroke timings can be same for many different key-pairs. 
Moreover, extrapolating information obtained from individual 
key-pairs to passwords presents a challenge that was only investigated with ad-hoc 
methods~\cite{halevi2015keyboard, compagno2017don}.

\subsection{Expected Contributions}
In this paper, we propose and evaluate a new human-based side-channel attack class, \attack,
based on thermal residue left behind by a user (victim) who enters a password using
a typical external keyboard. Shortly after password entry, the victim either steps away inadvertently, 
or is drawn away (perhaps as a result of being prompted by the adversary) from their personal workplace. 
Then, the adversary captures thermal images of the victim keyboard. We examine the efficacy of \attacka 
for a moderately sophisticated adversary equipped with a mid-range thermal imaging camera.  

To assess viability of \attacka, we conducted a rigorous two-stage user study. The first stage collected 
password entry data from 31 subjects using 4 common keyboards. In the second stage, 8 non-expert 
subjects acted as adversaries and 
attempted to derive the set of pressed keys from the thermal imaging data collected in the first stage. Our results show that 
even novice adversaries can use thermal residues to reliably determine the entire set of key-presses {\bf up to 30 seconds} 
after password entry. Furthermore, they can determine a partial set of key-presses as long as a full minute after password 
entry. We provide a thorough discussion of the implications of this study, and mitigation techniques against \attacka. Furthermore, in the course of exploring \attacka, we introduce a new post factum adversarial model. 

Due to inconsistencies in typing, we further find that thermal residue side-channel is not perfect, as it lacks 
information about password length, duplicate key-presses and key-press orderings. Inspired by these 
challenges, we utilize another (audio) side-channel within the same insider attacker model. This prompts
a new challenge in terms of how to combine these two side-channels. To this end, we design a general 
side-channel combination technique and describe a new \textit{hybrid attacker} model. We also introduce 
\combine attack which leverages both thermal residue and keyboard acoustics side-channels. This attack
closely corresponds to real-world insider attacks, i.e., no dictionaries -- which happens if  
random passwords are used, and no prior acoustic typing data of the victim. We evaluate this attack 
over numerous samples from 19 subjects entering representative passwords. Even with such limited capabilities, 
\combine greatly reduces the password search space.

\textbf{Organization.} Section~\ref{sec:background} gives the background for the paper. Section~\ref{sec:advers} introduces the adversarial models 
and Sections~\ref{sec:method} and~\ref{sec:rec} describe our methodology for exploiting individual side-channels 
and combination thereof. Section~\ref{sec:res} presents our results which is followed by discussions, 
related work and conclusion -- Sections~\ref{sec:dis},~\ref{sec:rw}, and~\ref{sec:conc}, respectively.

\section{Thermal Background}\label{sec:background}
\label{sec:bg}
This section provides some background on the physics of the thermal side-channel used   
in our experiments. Since keyboard acoustics have been extensively studied, we refer to~\cite{asonov2004keyboard} 
for a comprehensive discussion of keyboard acoustic emanations.

We start with a glossary of terms, then describe the form factor and material composition of the
modern 104-key ``Windows'' keyboards, and finish with certain Physics concepts. 
Given familiarity with elements of Conductive Heat Transfer and Newton's Law of Cooling, 
Sections~\ref{subsec:glossary},~\ref{subsec:conduct}, and~\ref{subsec:convect} can be skipped with 
no loss of continuity.

\subsection{Basic Thermal Terminology} 
\label{subsec:glossary}
\begin{compactitem}
\item Joule (J) -- Unit of energy Corresponding to $1$ Newton-Meter ($N \cdot {m}$)
\item Kelvin ($\degK$) -- Base unit of temperature in Physics. 
The temperature T in Kelvin (\degK) minus $273.15$ yields the corresponding temperature in degrees Celsius ($\degC$).
\item Watt (W) -- Unit of power corresponding to 1 Joule per second: ($\frac{J}{s}$)
\item Conduction --  Transfer of Thermal Energy caused by two objects in physical contact that are at different Temperatures.
\item Convection --  Transfer of Thermal Energy caused by submerging an object in a fluid.
\item Heat Transfer Coefficient - Property of a fluid that determines rate of convective heat flow. 
Expressed in Watts per square meter Kelvin: $\frac{W}{m^2 \degK}$
\item Specific Heat -- Amount of Thermal Energy in Joules that it takes to increase temperature of $1$kg 
of material by $1\degK$.  Expressed in Joules over kilograms degrees Kelvin: $J\over{kg}\degK$.
\item Thermal Conductivity --  Rate at which Thermal Energy passes through a material. Expressed in 
Watts per meters Kelvin: $W\over{m}\degK$
\item Thermal Energy -- Latent energy stored in an object due to heat flowing into it.
\item Thermal Source -- Object or material that can internally generate Thermal Energy such that it can stay at constant 
temperature during a thermal interaction, e.g., a heat pump.
\end{compactitem}

\subsection{Heating via Thermal Conduction}
\label{subsec:conduct}
Thermal Conduction is transfer of heat between any two touching objects of different temperatures. It is expressed as the 
movement of heat energy from the warmer to the cooler object. We are concerned with transfer of energy from a human 
fingertip to a pressed keycap. This transfer is governed by Fourier's Law of heat conduction which states that: 
\begin{quote}\em
Heat transfer between two objects can be modeled by the equation: 
$q = {\mathcal{K}A(T_1-T_2)t \over d}$, where $\mathcal{K}$ is thermal conductivity\footnote{$\mathcal{K}$ should not be
confused with $\degK$ -- degrees Kelvin.}
of the object being heated, $A$ is area of contact, $T_1$ is  initial temperature of the hotter object, $T_2$ is initial temperature of 
the cooler object,  $t$ is time, and $d$ is the thickness of the object being heated.  
\end{quote}
The relationship between an object's heat energy and its temperature is governed by the object's mass and 
specific heat, as dictated by the formula: $q = c m \Delta T$, where $q$ is total heat energy, $c$ is object's specific heat, 
$m$ is object's mass and $\Delta T$ is change in temperature.

We consider the human body to be a thermal source, and we assume that any change in the fingertip temperature 
during the (very short) fingertip-keycap contact period is negligible, due to internal heat 
regulation~\cite{dai2004comparison}. Furthermore, we assume that:
\begin{compactitem}
\item Average human skin temperature is $307.15\degK$ ($= 34\degC$)~\cite{burton1939range}.
\item Keyboard temperature is the same of that as that of the air, which, for a typical office, 
is OSHA\footnote{OSHA = Occupational Safety and 
Hazards Administration, a United States federal agency.}-recommended
$294.15\degK$ ($=21\degC$)~\cite{occupational1999osha}.
\item Keycap area is 0.00024025 $m^2$, keycap thickness is 0.0015 meter and keycap 
mass is $.4716g$ (See: Section~\ref{subsec:composition}).
\item Average duration of a key-press is $0.28$s~\cite{sauro2009estimating}.
\end{compactitem}
Therefore, for variables mentioned above, we have: 

\begin{center}
\fbox{
 \bf\footnotesize

$\mathcal{K}$=0.25, $\;$ A=0.00024025, $\;$ T1=34, $\;$ T2=21, $\;$  t=0.28, $\;$ and $\;$ d=0.0015 }
\end{center}

Plugging these values into Fourier's Law, we get: 
\begin{equation}
q = {(0.25)(0.00024025)(34-21)(.28) \over 0.0015}
\end{equation}
which yields 
total energy transfer: $q=0.1458$J.  We then use total energy $q$ in the specific heat equation to determine total temperature 
change: $0.1458=(1000)(0.0004716)\Delta T$. This gives us a total temperature change of $\Delta T = 0.3092$. 
Therefore, we conclude that the average human fingertip touching a keycap 
at the average room temperature results in the keycap heating up by $0.3092\degK$.

\subsection{Cooling via Thermal Convection}
\label{subsec:convect}
After a keycap heats up as a result of conduction caused by a press by a warm(er) human finger, it begins to cool off 
due to convective heat transfer with the air in the room. Convection is defined as the transfer of heat resulting from the 
internal current of a fluid, which moves hot (and less dense) particles upward, and cold (and denser) particles -- downward. 
This interaction is governed by Newton's Law of Cooling. Its particulars are impacted by the shape and position of the 
heated object. In our case, there is a plane surface\footnote{The actual keycap surface can be slightly concave.} 
facing towards the cooling fluid (i.e., a keycap directly exposed to ambient air)
which is described by the formula: 
\begin{equation}
    T(t) = T_{s} + (T_{0} -T_{s})e^{-\kappa t}
\end{equation}

where $T(t)$ is temperature at time $t$, $T_s$ is temperature of ambient air, 
$T_0$ is initial object temperature, and $\kappa$ is the cooling constant of still (non-turbulent) air 
over a $0.00024025m^2$ plane.

This comes with the additional intuitive notion that a surface convectively cools quicker when the 
temperature difference between the heated object and the fluid is higher. Similarly, it cools slower when 
the temperature difference is smaller. Finally, Newton's Law of Cooling is asymptotic, and cannot be used 
to find the time at which the object reaches the exact temperature of the ambient fluid. 
Thus, instead of finding the time when the temperatures are equal, we determine the time when the temperature 
difference falls below an acceptable threshold, which we set at $0.04\degK$. Plugging this into Newton's 
Law of Cooling results in: 
\begin{equation}
t = - {ln({0.3092\over 0.04}) \over 0.037}
\end{equation}
which yields $t= 55.7$ for total time for a pressed key to cool down to the point where it is indistinguishable from the 
room temperature.

\subsection{Modern Keyboards}
\label{subsec:composition}
\begin{figure}[ht!]
	\begin{minipage}{.5\linewidth}
		\centering\fbox{
			\includegraphics[height=1.2in]{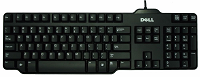} }
		\caption{\small{Typical ``Windows''-style Keyboard.}}
		\label{fig:board}
	\end{minipage}
	\hspace{2cm}
	\begin{minipage}{.3\linewidth}
		\centering\fbox{
			\includegraphics[height=1in]{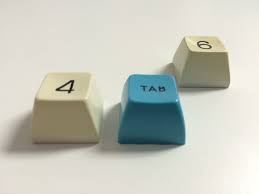} }
		\caption{\small{Typical Keycap Profile.}}
		\label{fig:keycap}
	\end{minipage}
\end{figure}

Most commodity external keyboard models are of the 104-key ``Windows'' variety, shown in Figure~\ref{fig:board}. 
On such keyboards, the distance between centers of adjacent keys is
about $19.05$mm, and a typical keycap shape is an  $\approx[15.5mm~x~15.5mm~x~1.5mm]$ rectangular prism, with 
an average travel distance of $3.55mm$~\cite{noyes1983qwerty}; see Figure~\ref{fig:keycap}. 
All such keyboards are constructed out of Polybutylene Terephthalate (PBT) with density of $1.31g/cm^{3}$ , 
resulting in an average keycap mass of $.4716g$~\cite{pyda2004heat}.  PBT generally has the following characteristics: 
specific heat = $1,000{J \over kg \degK}$ and thermal conductivity = $0.274{W\over m\degK}$~\cite{pyda2004heat}.

\subsection{Thermal Cameras}
\label{subsec:FLIRS}
\begin{figure}[ht!]
	\fbox{\centering
		\includegraphics[height=1.7in,width=.75\columnwidth]{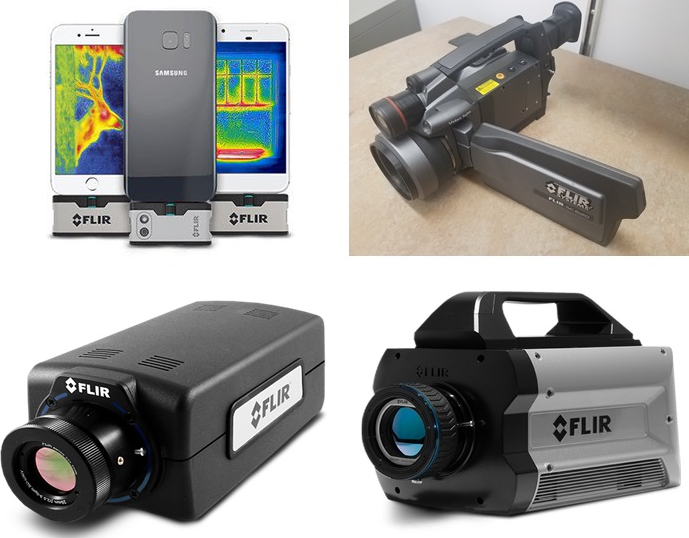} }
	\caption{\small{FLIR Devices / Thermal Imagers: FLIR ONE(top left), SC620 (top right), 
	A6700sc (bottom left) and X8500sc (bottom right).}}
	\label{fig:flirs}
\end{figure}

In the past few years, many niche computational and sensing devices have moved from Hollywood-style fantasy
into reality. This includes thermal imagers or cameras. In order to clarify their availability to individuals (or agencies) 
at different levels of sophistication, we provide the following brief comparison of several types of readily-available 
FLIR: {\bf F}orward-{\bf L}ooking {\bf I}nfra-{\bf R}ed devices. (See: Figure~\ref{fig:flirs} for product images 
and \url{https://www.flir.com/products} for full product specifications.)
In the rest of the paper, we use the following terms interchangeably: FLIR device, thermal imager and thermal camera.
\begin{compactenum}
\item[FLIR One] -- Price: About US\$$300$. Thermal Sensitivity: $0.15$K. Thermal Accuracy:  
$\pm 1.5$K or $1.5\%$ of reading. Resolution:$50x80$. Image Capture: Manual, $1$ image at a time. Video Capture: None 
\item[SC620] -- Price: About US\$$1500$ (used). Thermal Sensitivity: $0.04$K Thermal Accuracy: 
$\pm 2$K or $2\%$ of reading. Resolution: $640x480$. Image Capture: Automatic, programmable to capture 
images by timer, or when specific criteria are met, at maximum rate of $1$ image per second. Video Capture: None.
\item[A6700sc] -- Price: About US\$$25,000$. Thermal Sensitivity: $0.018$K Thermal Accuracy: 
$\pm 2$K or $2\%$ of reading. Resolution: $640x512$. Image Capture: Automatic, programmable to capture  
images by timer or when specific criteria are met, at up to $100$fps. Video Capture: High speed, up to 
$100$fps.
\item[X8500sc] -- Price: About US\$$100,000$. Thermal Sensitivity: $0.02$K: Thermal Accuracy: 
$\pm 2$K  or $2\%$ of reading. Resolution: $1280x1024$ Image Capture: Automatic, programmable to capture 
images by timer or when specific criteria are met, at up to $180$fps. Video Capture: High speed, up to 
$180$fps.
\end{compactenum}
Obviously, a sufficiently motivated organization or a nation-state could easily obtain thermal 
imagers of the highest quality and price. However, we assume that the anticipated adversary 
is of a mid-range  sophistication level, i.e., capable of acquiring a device exemplified by SC620.
However, we note the adversary armed with a FLIR One (which is on the low-end of the spectrum for thermal imagers,
and can be connected to any commodity smartphone  without substantially altering the overall form factor) can collect 
thermal residues up to $20$ seconds after entry.  Whereas, the adversary with a A6700sc or X8500sc can do the same
$139$ seconds, and $136$ seconds after entry, respectively. Also, since thermal residues decay at a logarithmic rate, 
future advances in thermal camera sensitivity will result in a exponential increase of collection time. 
\section{Adversarial Model \& Attacks}
\label{sec:advers}
This section describes the adversarial models for \system and \acutherma and defines \hybrida.

\subsection{\attack Attack}
Fourier's Law states that contact between any two objects with unequal temperatures 
results in transfer of heat energy from the hotter to the cooler object. It is reasonable to assume that the typical office 
environment has the ambient temperature within the OSHA-recommended range of $293.15-298.15\degK$ 
(=$20-25\degC$)~\cite{occupational1999osha}. In that setting, the average human hand is expected to conductively 
transfer an observable amount of heat to the ambient-temperature keyboard. Consequently, a bare-fingered human 
typist cannot avoid leaving thermal residue on a keyboard. 
This physical interaction can be exploited by the adversary in order to harvest the thermal residue of a victim who recently 
used a keyboard to enter potentially sensitive information, e.g., a password. This forms the premise for \attacka.

\subsubsection{Attack Scenario}
\begin{figure}[!ht]
	\centering
	\begin{subfigure}{0.3\linewidth}
		\includegraphics[width=\textwidth]{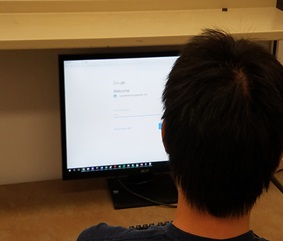}
		\caption{STEP 1: Victim Enters Password} 
	\end{subfigure}
	\hspace{0.2cm}
	\begin{subfigure}{0.35\linewidth}
		\centering
		\includegraphics[height=0.9in, width=.7\textwidth]{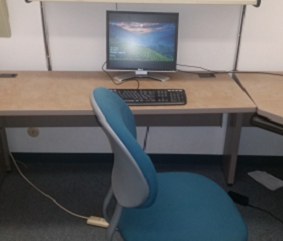}
		\caption{STEP 2: Victim Leaves \\(\textit{Opportunistic})}
		\vspace{0.4cm}
		\includegraphics[height=0.9in, width=.7\textwidth]{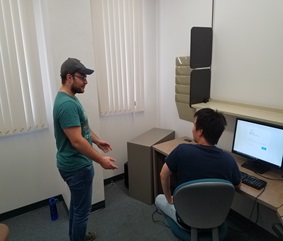}
		\setcounter{subfigure}{1}
		\caption{STEP 2: Victim Drawn Away \\(\textit{Orchestrated})}
	\end{subfigure}
	\begin{subfigure}{0.3\linewidth} 
		\includegraphics[width=\textwidth]{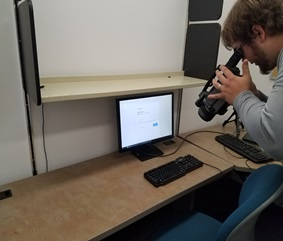}
		\caption{STEP 3: Thermal Residues Captured} 
	\end{subfigure}
	\caption{An Example of \attack Attack.} 
	\label{attack}
\end{figure}

\attack is a type of insider attack, which proceeds as follows 
(see Figure~\ref{attack}):
\begin{description}
\item[STEP 1:] The victim uses a keyboard to enter a genuine password, as part of the log-in (or session unlock) procedure.
\item[STEP 2:] Shortly thereafter, the victim either: willingly steps away, or is lured away, from the workplace.
\item[STEP 3:] Using thermal imaging (e.g., photos taken by a commodity thermal camera) the adversary 
harvests thermal residues from the keyboard.
\item[STEP 4:] At a later time, the adversary uses the ``heat map'' of the images to determine recently pressed keys. 
This can be done manually (i.e., via visual inspection) or automatically (i.e., via specialized software). 
\item [\underline{REPEAT:}] The adversary can choose to repeat STEPS [1-4] over multiple sessions. 
\end{description}

The two options in \textbf{STEP 2} correspond to \emph{opportunistic} and \emph{orchestrated} attack sub-types, 
respectively. In the former, the adversary patiently waits for the opportunity: once the victim leaves 
(on their own volition) shortly after password entry, the adversary swoops in and collects thermal residues. 
This strategy is similar to \luncha. In an \emph{orchestrated} attack, instead of waiting for the victim to leave, 
the adversary uses an accomplice to draw the victim away shortly after password entry.

\changed{We argue that either of these attack scenarios, which are in line with previous literature exploiting thermal camera emanations (\cite{thermanator,SafeCracking,andriotis2013pilot, 
abdelrahman2017stay,mowery2011heat, cardaioli2020your}), is very plausible. For a typical workday, employees spend a significant portion of their time in meetings, phone calls, and other non-core tasks, often switching between them~\cite{leroy2009so}. 
Each of these activities may require them to leave their assigned workspace, and most of the time, users (voluntarily) leave their logged-in sessions (and workstations) unattended, opening the risk to a plethora of attacks~\cite{conti2020auth,cardaioli2022privacy}. 
Therefore, such unplanned tasks may appear shortly after a password entry since they may be unplanned or in the spur of the moment~\cite{thermanator}.
Similarly, many employees go to the restroom anytime they want while at work~\cite{hartigan2021real}, e.g., after the login phase while the PC executes the startup processes or applications, which can take a while before it becomes fully functional. Indeed, some applications, such as antivirus, instant-messaging platforms, or cloud-synchronizations, could take a while to become operative, making the PC slow (or unusable) for seconds or even minutes~\cite{slowstart}. During such a period, people may also go and say hello to colleagues or have coffee. 
Considering insider attacks' long-term nature, an attacker will eventually observe such a situation and will mount the \attack attack.

As mentioned before, in addition to job-related tasks, social gatherings at work present additional opportunities for \attack attackers. 
For instance, coffee breaks are when ``employees come together spontaneously and thus contribute to ``natural internal fluid information circulation''~\cite{barmeyer2019informal} (note the spontaneity).
In fact, coffee breaks are considered to be of utmost importance for employees' well-being in certain professions, and phrases such as ``Let's Have a Cup of Coffee!'' are equivalent to having a conversation~\cite{topik2009coffee}.
Such social gatherings could be instigated by accomplices right after a password entry, and given their importance, the victims are likely to comply.

Furthermore, some professions may offer more opportunities for such attacks. For instance, in healthcare, nurses often have short~\cite{bowers2001nurses} conversations with patients before calling in the doctor during which they might need to log in to the system using their password.
This might be due to recording the condition the patient is in (e.g., complaints, blood pressure) or confirming patient identity. Afterwards, nurses may log out and leave the room.
A malicious patient could then take the thermal pictures of the keyboard thereafter and potentially gain access to many other patients' records using the recovered password.

Although \attack attack relies on a dedicated device (i.e., a thermal camera), such cameras have similar designs to video cameras (as is the case for SC620 used in our experiments) which will avoid suspicion of an attack.
Furthermore, a dangerous aspect of insider attacks is that insider attackers could wait for the perfect opportunity to mount their attacks.
Owing to this, an attacker can wait for no one to be around for mounting a \attack attack or could employ an accomplice to make sure the attack cannot be observed by others.
Moreover, as the quality of thermal imagers are improving -- resulting in clearer images, their size is decreasing.
It is even possible to attach a small thermal imager to a mobile phone making the attack virtually undetectable.
We present initial results with this camera in Section~\ref{sec:conc}.
}

\subsection{Hybrid Attacks}
%
\hybrida are a type of insider attack that use multiple side-channels. 
It can be mounted on multiple devices and systems using several side-channels. Depending on the exact
side-channels, the attacker might exploit individual side-channels  simultaneously, or at different times. 
This attack model is particularly useful since an insider attacker falls into the ``covert adversary'' model. A covert adversary, 
as defined in~\cite{aumann2007security}, can cheat (we define cheating as trying to obtain a secret) yet does not want to get 
caught. Using unsecured and/or overlooked side-channels, the adversary increases his chance of avoiding detection. On the
other hand, a side-channel might provide less information about the target secret, as compared to a more direct channel. 
Thus, by combining multiple side-channels, the attacker increases his chance of gaining more information 
on the target secret.

\subsection{\acutherm\ Attack}
\acutherm\ targets passwords entered using external keyboards, using both 
thermal and acoustic side-channels, under the following assumptions:

\begin{assumption}\label{asmp1}
	The attacker has a mid-range thermal camera and a commodity 
	recording device, i.e., a microphone.
\end{assumption}

\begin{assumption}\label{asmp2}
	The attacker has physical access to the victim's keyboard, though no physical contact 
	between the attacker and the keyboard is needed.
\end{assumption}

\begin{assumption}\label{asmp3}
	The attacker knows whether the victim is in her workspace and can influence the victim's 
	presence in that workspace.
\end{assumption}

\begin{assumption}\label{asmp4}
	The attacker knows when the victim is entering a password. 
\end{assumption}

Assumption~\ref{asmp1} captures the equipment requirement for \acutherm: the total cost of the attack is rather low, e.g., 
\$1,500 for a mid-range thermal camera (FLIR SC620) + \$50 for a higher-end microphone.
As an insider (e.g., an office-mate of the victim), the attacker satisfies Assumptions~\ref{asmp2} and~\ref{asmp3}. 
Assumption~\ref{asmp2} lets the attacker build a keyboard acoustic profile of the victim's keyboard  
(\textbf{but not of the victim's typing!}), record the sounds produced during password entry, and 
take a thermal picture soon thereafter. Assumption~\ref{asmp3} is related to stealthiness. 
Combined with Assumption~\ref{asmp2}, it allows the attacker to take a thermal image of the 
keyboard soon after password entry either by waiting for the victim to leave on their own accord, 
or by having an accomplice lure the victim away. Assumption~\ref{asmp4} is satisfied by the fact that, if the victim's workstation is locked or not logged in, the first thing the victim normally enters is the  password. 

\subsubsection{Attack Scenario}
\begin{figure}[!h]
	\centering
	\begin{minipage}{.35\linewidth}
			\includegraphics[width=\linewidth]{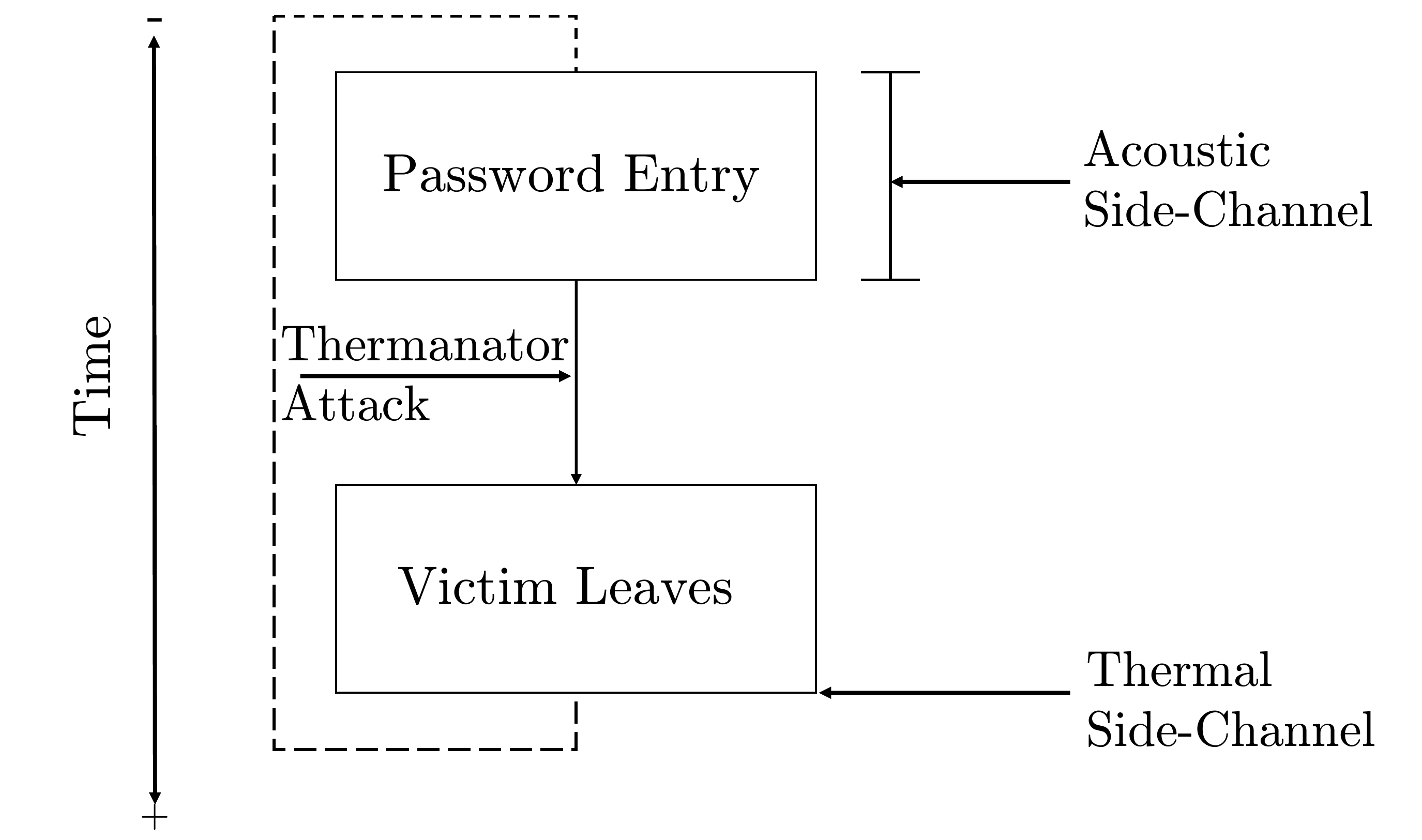}
		\caption{ \acutherm\  timeline: thermal and acoustic side-channels can be used independent 
			of each other and multiple times. Dashed lines represent another attack instance.}
		\label{tkz:att}
	\end{minipage}
	\hspace{0.2cm}
	\begin{minipage}{.60\linewidth}
		\includegraphics[width=\linewidth]{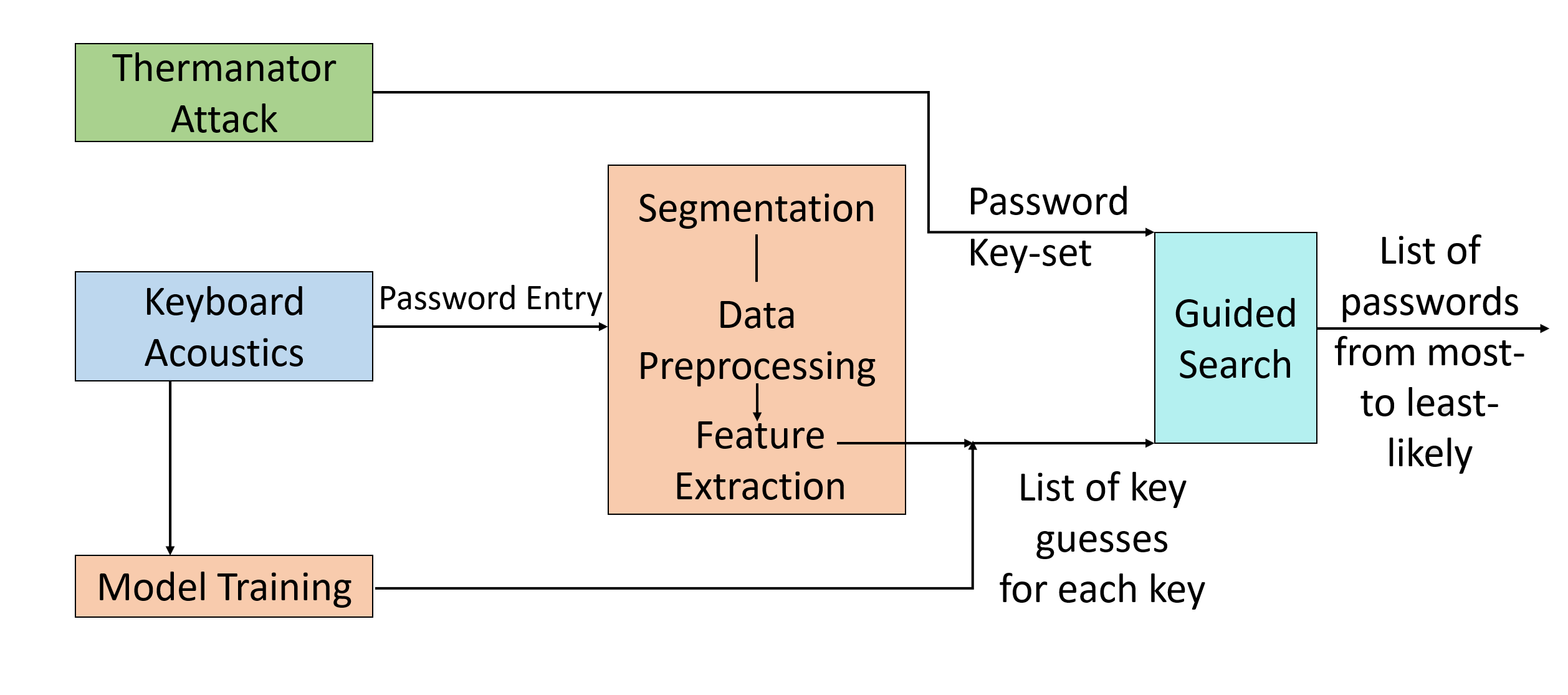}
		\caption{ \acutherm\ Attack Overview.}
		\label{fig:overview}
	\end{minipage}
\end{figure}

In the offline phase, the attacker trains a machine learning model on the sounds produced by key-presses on the
victim's keyboard. Since building a model of the victim's typing might not be feasible, the training dataset consists 
of acoustic emanations of key-presses of the attacker. We expand on how this can be achieved for various typing 
styles in Section~\ref{sec:prof}. During password entry, the attacker records the sounds produced by the victim 
typing the password. Afterwards, the attacker takes a thermal image of the victim's keyboard, as in  
the \attacka. As mentioned earlier, this attack can be \textit{opportunistic} or \textit{orchestrated}. 
(See Figure~\ref{tkz:att} for attack timings.)

Thermal image analysis consists of the attacker determining the keyboard regions with higher temperatures 
(i.e., ``hot'' or ``lit'') on the thermal image. This step yields a set of candidate keys. To use keyboard acoustic 
emanations, the attacker applies the pre-trained model to obtain a list of guesses for each key, after performing 
pre-processing, segmentation and feature extraction on recorded sounds. Then, \textit{guided search} is used to 
rank candidate passwords. We describe this step in detail in Section~\ref{sec:combine}. A visual representation of the 
attack is in Figure~\ref{fig:overview}.

\section{Materials and Methods}
\label{sec:method}
This section describes the experimental apparatus, procedures, and subject recruitment methods.

\subsection{Apparatus}
\label{subsec:app}
\begin{figure}[th!]
\centering 
\begin{minipage}{0.45\linewidth}
	\centering
	\includegraphics[height=1.3in]{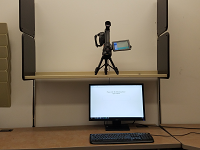} 
	\caption{\small{SC620 Apparatus Setup}}
	\label{fig:620Setup}
\end{minipage}
\begin{minipage}{0.45\linewidth}
\centering
	\includegraphics[height=1.3in]{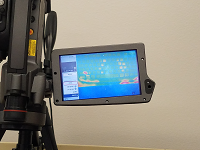} 
	\caption{\small{Example of Thermal Emanations being Recorded.} }
	\label{fig:620Data}
\end{minipage}
\end{figure}

\begin{figure}[ht!]
	\centering
	\begin{subfigure}{0.22\linewidth}
		\includegraphics[width=\textwidth]{keyboard.png}
		\caption{Dell SK-8115} 
	\end{subfigure}
	\begin{subfigure}{0.22\linewidth}
		\includegraphics[width=\textwidth]{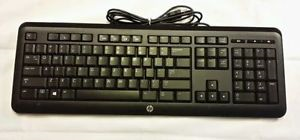}
		\caption{HP SK-2023} 
	\end{subfigure}
	\begin{subfigure}{0.22\linewidth} 
		\includegraphics[width=\textwidth]{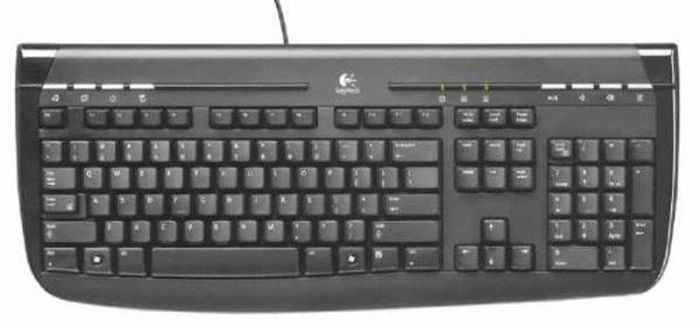}
		\caption{Logitech Y-UM76A.} 
	\end{subfigure}
	\begin{subfigure}{0.22\linewidth} 
		\includegraphics[width=\textwidth]{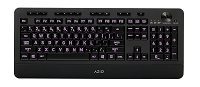}
		\caption{AZiO Prism KB507 (backlit).} 
	\end{subfigure}
	\caption{Keyboards used in our experiments.} 
	\label{kboards}
\end{figure}

The experimental setup was designed to simulate a typical office setting. It was located in a dedicated 
office in a research building of a large university. Since experiments were conducted during the academic year,
there was always some (though not excessive) amount of typical office-like ambient noise. Figure~\ref{fig:620Setup} 
shows the setup from the subject's perspective, while Figure~\ref{fig:620Data} shows an example of Thermal Emanations being recorded.
Equipment used in \attack experiments consisted of the following readily available (off-the-shelf) components:
\begin{compactenum}
\item[I.] FLIR Systems SC620 Thermal Imaging Camera\footnote{see: \url{http://www.FLIR.com} for a full specification.} 
This camera was perched on a tripod $24\,''$ above the keyboard.
\item[II.] Four popular and inexpensive commodity computer keyboards (See Figure~\ref{kboards}):
(a) Dell SK-8115,
(b) HP SK-2023
(c) Logitech Y-UM76A, and
(d) AZiO Prism KB507.
\end{compactenum}

In \hybrid experiments, we used the same equipment, except for removing the HP keyboard and
introducing a Yeti professional microphone made by Blue \footnote{See: 
https://www.bluedesigns.com/products/yeti/ for a full specification}.
The particular thermal camera that was used in our experiments 
was chosen to be realistic for a moderately sophisticated and determined adversary. We assume
this type of adversary to be an individual, i.e., not an intelligence agency, a nation-state, or a powerful criminal organization. 
FLIR SC620 Thermal Imager costs approximately $\mbox{US}\$1,500$ used. (This model is about 6-7 years old.) 
It automatically records images at the resolution of $640x480$ pixels, with $1Hz$ frequency. Its thermal sensitivity is $0.04\degK$. 

The four keyboards were chosen to cover the typical range of manufacturers represented in an average workplace. 
Dell, HP and Logitech keyboards are popular default keyboards included in new computer orders from major 
PC, desktop, and workstation manufacturers. Each costs $\approx\mbox{US}\$20$. 
Meanwhile, Azio Prism is a popular low-cost and independently manufactured keyboard that can be 
easily obtained on-line e.g., from Amazon for $\approx\mbox{US}\$25$.

\subsection{Subject Recruitment Procedure}
Subjects were recruited from the student body of a large university using a unified Human Subjects Pool 
designated for undergraduate volunteers seeking to participate in studies such as ours. Subjects were compensated 
with course credit. Unsurprisingly, the overwhelming majority of subjects were in the $18-25$ age range. 
We collected data from 31 ($16$ male and $15$ female) and 19 ($11$ male and $8$ female) people for the \attack and \hybrid attacks respectively.

All experiments were duly authorized by the Institutional Review Board (IRB) of the authors' employer, 
well ahead of the commencement of the study. The level of review was: Exempt, Category II. 
No sensitive data was collected during the experiments and minimal identifying information was retained. 
In particular, no subject names, phone numbers or other personally identifying  information (PII) was collected.
All data was (and is) stored pseudonymously.

\subsection{\attack Procedures}
\label{subsec:proc}
\system was evaluated using a two-stage user study. The first stage was conducted to collect thermal 
emanation data, and the second -- to evaluate efficacy of \attacka. A given subject only participated in a 
single stage.

\subsubsection{Stage One: Password Entry}
\begin{figure}[ht!]
\centering
\begin{minipage}{.7\linewidth}
		\includegraphics[width=\textwidth]{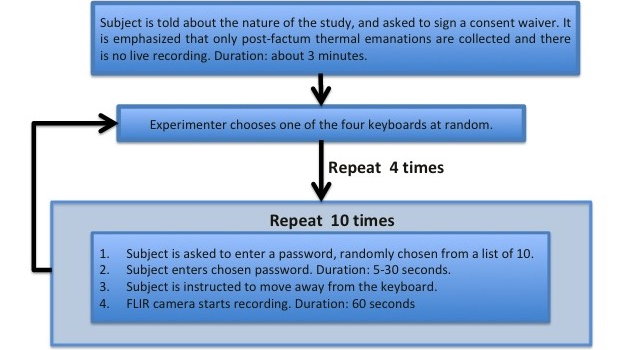}
	\caption{\small{Experiment Stage One: Flowchart}}
	\label{fig:flowchart}
\end{minipage}
\hspace{-2ex}
\begin{minipage}{.28\linewidth}
		\includegraphics[width=\textwidth]{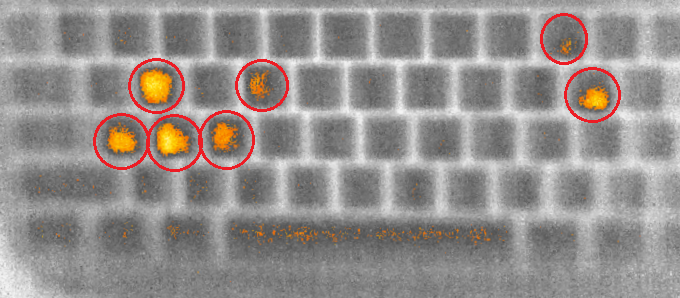} 
		\caption{\small{Thermal image of ``passw0rd'' $20$ seconds after entry.}}
		\label{fig:pw20}
\end{minipage}

\end{figure}

Recall that \attack 's goal is to capture thermal residues of subjects {\bf after} keyboard password entry.
This is accomplished by having FLIR SC620 take a sequence of images (60 total), one per second, for a total
of one minute after initial password entry. The first stage is shown in Figure~\ref{fig:flowchart}. 
This collection of 60 images does {\bf not} represent the requirements for a single attack. 
In reality, the adversary would arrive as quickly as possible (after the victim leaves the workspace) and take a 
single thermal image. For strictly experimental purposes, a full minute of thermal data was captured to more 
accurately model adversaries arriving after some time has elapsed. 

Each subject entered $10$ passwords on $4$ keyboards and each entry was followed by one minute of keyboard 
recording ($60$ successive images) by the FLIR. Each subject entered a total of $40$ passwords and every entry took, 
on average, between $10$ and $20$ seconds. The total duration of the experiment for a Stage One subject ranged 
between $50$ and $60$ minutes, based on the individual's typing speed and style. 
Both keyboards and passwords were presented to each subject in random order, in an attempt to negate
any side-effects due to subject training or familiarity with the task. 

We selected 10 passwords that included both ``insecure'' and ``secure'' categories. The former were culled from 
the top 100 passwords by popularity that adhere to common password requirements, such as Gmail 
\footnote{see: \url{https://support.google.com} for details}. Whereas, ``secure'' passwords were created by randomly 
generating 8-, 10-, and 12-character strings of lower/uppercase letters as well as numbers and symbols 
that adhere to Gmail restrictions. Our selection criteria resulted in the following 10 candidate passwords: 
\begin{compactitem}
\item {\bf [Insecure]:} ``password'', ``12345678'', ``football'', ``iloveyou'', ``12341234'', \\``passw0rd'', and ``jordan23'',
\item {\bf [Secure]:} ``jxM\#1CT['', ``3xZFkMMv$|$Y'', and ``6pl;0$>$6t(OvF''.
\end{compactitem}

\subsubsection{Stage Two: Data Inspection}
The second stage of the experiment had subjects, who act as adversaries. 
Subjects were shown images obtained from the first stage of the experiment, e.g., Figure~\ref{fig:pw20}, 
and were instructed to identify the ``lit'' regions. Each subject was shown $150$ recordings of password entries 
in random order. On average, a subject could process a single recording in $45-60$ seconds. Total time for each 
Stage Two subject was around $100-130$ minutes.

\subsection{\hybrid Procedures}
Recall that we aimed to collect both thermal and acoustic emanations. While capturing thermal residue is 
the same as in \attack,  we need acoustic emanations to compensate for the thermal side-channel's lack of 
password length, duplicate key-presses and key-press ordering. We used the Yeti microphone to record 
password entry. Each subject entered 10 randomly displayed passwords. Passwords were presented in 
random order to mitigate any effect of expected ordering on subject performance. This process was repeated 
three times, once for each keyboard. Allotting between 10 to 20 seconds per password, the total duration of 
the experiment for a single subject ranged between 40 and 50 minutes, based on one's typing speed and style. 
We used the same 10 passwords as in \attack.

\changed{Due to subjects' prior unfamiliarity with (and random nature of) secure passwords, they experienced some 
difficulties entering these passwords. They often pressed wrong keys and cleared them with backspace, 
rendering the resulting audio useless for our purposes. In addition, for the same reason, timing information 
obtained from secure passwords did not reflect a natural password entry. This is because a real-world user
who has a secure password is a-priori familiar with it and thus enters it faster than an unfamiliar random password.

Moreover, the capital letters in the ``secure'' passwords introduced additional challenges in our dataset. First, the detection of the Shift key is different from all the other keys. This is due to the release of Shift occurring after release of the corresponding letter or symbol, so its complete key sound (press + release) would not appear clearly in the recordings. Second, both Left and Right Shift can be used for capital letters, but they present different acoustic features, complicating the detection. Third, to enter capital letters, some users used Shift + Letter combination, others used Caps Lock even for a single letter, and few participants mixed the two techniques, making it harder to determine if and when capital letters were used. For these reasons, 
``secure'' passwords used in \attack are not suitable for this first stage of \hybrid attack experiments and 
were not included in our analysis. ``Insecure'' passwords, not containing capital letters, were not affected by this limitation. We plan to focus and implement upper case detection in future work.} 

However, we believe that our results generalize to ``secure'' passwords since: \textit{(1)} We do not employ dictionaries 
to reduce the search space, since this would be impossible for random passwords, and \textit{(2)} The typing behavior for 
``secure'' passwords approximates that of ``insecure'' passwords with enough repetition, which is expected. Since even 
though these passwords are random, they are typed many times by their owners. Although we would have preferred to use 
subjects' ``secure'' passwords in our experiments, this was not possible due to our IRB.

Note that, in some cases, there were spurious or extraneous sounds coming from the environment and for some passwords. 
Thus, we had to remove some password instances from the evaluation set. This is within attacker's capabilities 
since one can record password entry multiple times until a good sample is obtained; this is due to the opportunistic 
nature of  \hybrid attacks. For some characters in passwords, the tool we used from~\cite{stcode} did not produce predictions due to an error we could not fully track to its origin. We excluded such passwords from our password results (e.g., password space reduction) but included them in general character-based results (e.g., reduction per-keyboard). We provide a list and frequency information of all passwords used in our experiments in 
Section~\ref{sec:res}.

\subsection{Acoustic Side-Channel Exploitation}
To use the acoustic side-channel information, we pre-processed the recorded audio and extracted the features 
to build a model capable of recognizing pressed keys.

\subsubsection{Pre-processing}

To extract pressed keys from the raw signal, we need to detect when key-presses occur and then split the audio 
accordingly. To this end, we first filter out the signal with a Butterworth Filter and keep the frequencies between 400 and 
12KHz to remove noise. These frequencies were shown to be relevant in~\cite{zhuang2009keyboard}. 
Then, we follow an approach similar to~\cite{berger2006dictionary}. We fragment the signal into windows of $88$ 
samples, which corresponds to $2$ms each.  We then apply the Fast Fourier Transform (FFT) to the extracted signal, 
and, by summing respective FFT coefficients, calculate an indication of the amount of energy in each window. 
We then create an energy vector and normalize it with values between $0$ and $1$. Going through the energy vector 
and calculating the difference between each window and its predecessor, we find the positive energy change 
for each window. If this change exceeds a certain threshold, the window is expected to contain a key-press. 

Once a key-press is detected, we look for another key-press, skipping a fixed amount of time (usually $100-150$ms), 
which represents the probable interval within which two key presses cannot coexist. The energy threshold and the 
skipped time have to be tuned for each person, since the typing style (e.g., key pressing speed and pressure) 
differs among individuals. To find the key release, we look for an energy increase between two key-presses. 
If the energy is lower than a threshold related to its key-press, it is ignored. \changed{In the experiments, human visual inspection assisted the automatic key-press and release detections. Visual inspection is reasonable from the attacker's point of view. It can increase the detection accuracy and is not time-consuming, considering he would attack a single password per time. Indeed, it is not challenging to supervise the keys detection algorithm by hearing at the password recording and looking at its raw signal.}

\begin{figure}[ht!]
	\includegraphics[width=\linewidth, height=4cm]{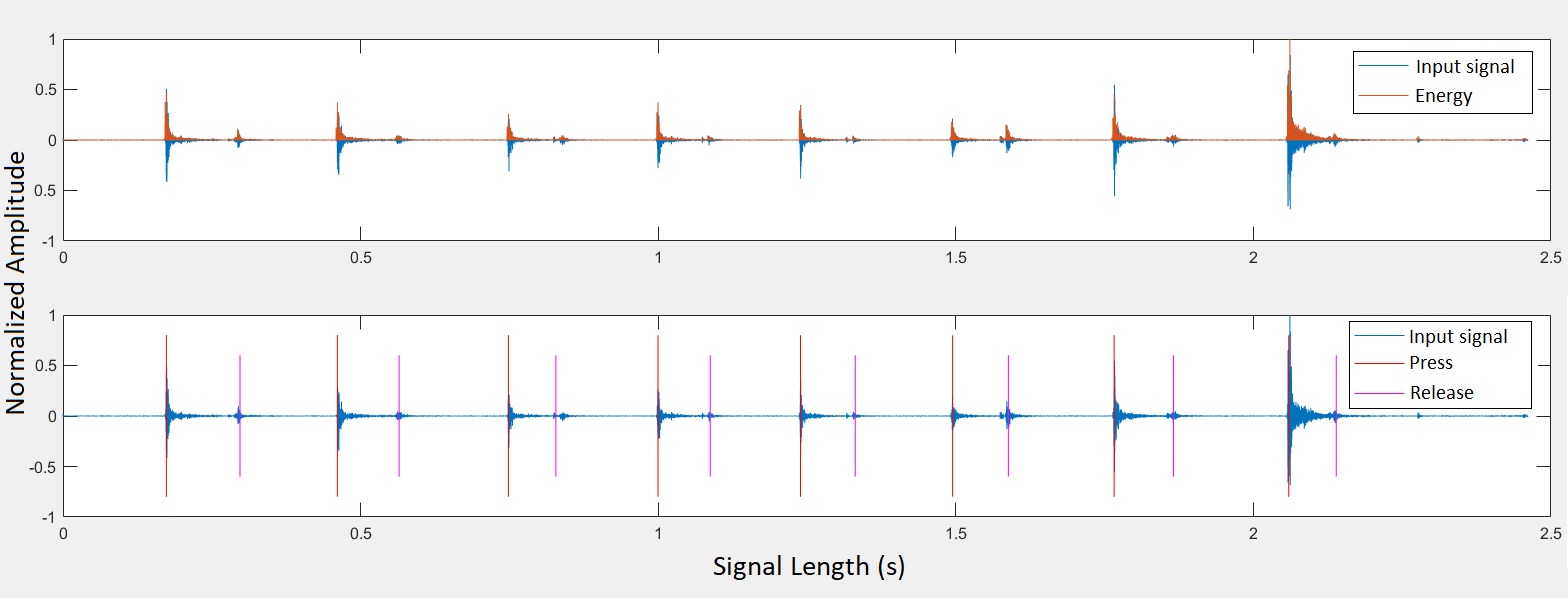}
	\caption{Press energies (top) and press and release times (bottom). }
	\label{fig:press}
\end{figure}

After all key-presses are found, we split the audio using the time of the starting points. Figure~\ref{fig:press} (top) 
shows the original signal (blue) and the related energy (orange). Figure~\ref{fig:press} (bottom) shows the 
times of key presses (red) and releases (purple) after the process. 

\subsubsection{Feature Extraction}
We use the same features as in~\cite{compagno2017don}: mel-frequency cepstral coefficients (MFCC) and the 
code available from~\cite{stcode} for identifying keystrokes. 
MFCC features have been effectively used in speech/speaker recognition~\cite{ittichaichareon2012speech,tiwari2010mfcc} 
and provide information on characteristics of speech/signal. 
Similar to~\cite{compagno2017don}, we use a sliding window of 10ms with a step 
size of 2.5ms, 32 filters in the mel scale filterbank, and use the first 32 MFCC since it was shown that these features 
were optimal in conjunction with MFCC.


\subsubsection{Profiling}\label{sec:prof}
Our model (See Section~\ref{sec:advers}) assumes that the attacker has access to the victim's keyboard, but does not have an acoustic profile of the victim's typing on that keyboard. The sophistication level of this adversary is lower than the complete profiling scenario~\cite{compagno2017don}, where the attacker trains a model on a victim's actual typing, which yields significantly better key accuracies.

\changed{To recognize a pressed key from its MFCC features, we train a machine learning model to solve a multi-classification task (one class for each key). The model outputs the predicted key along with other possible keys in descending probability.}
To train a model, we use the same keyboard as the victim uses to enter passwords. Considering that keystroke characteristics depend on one's typing style~\cite{halevi2015keyboard}, we train Hunt-and-Peck and Touch Typing models (HP and TT, respectively) by entering each of the 46 characters 10 times on a commodity 101-key keyboard, using the corresponding typing style. These characters include: $26$ letters in the English alphabet, 10 digits $\{0-9\}$ and $10$ symbols \{.=-,;'[]/$\backslash$\}. Also, to reduce the effects of different typing styles, we 
combine these models and create an additional model based on both HP and TT, which we refer to as HPTT. This model is trained on 10 keystrokes of 46 characters entered using each typing style. 

\begin{figure}
\centering
\includegraphics[width=0.9\linewidth]{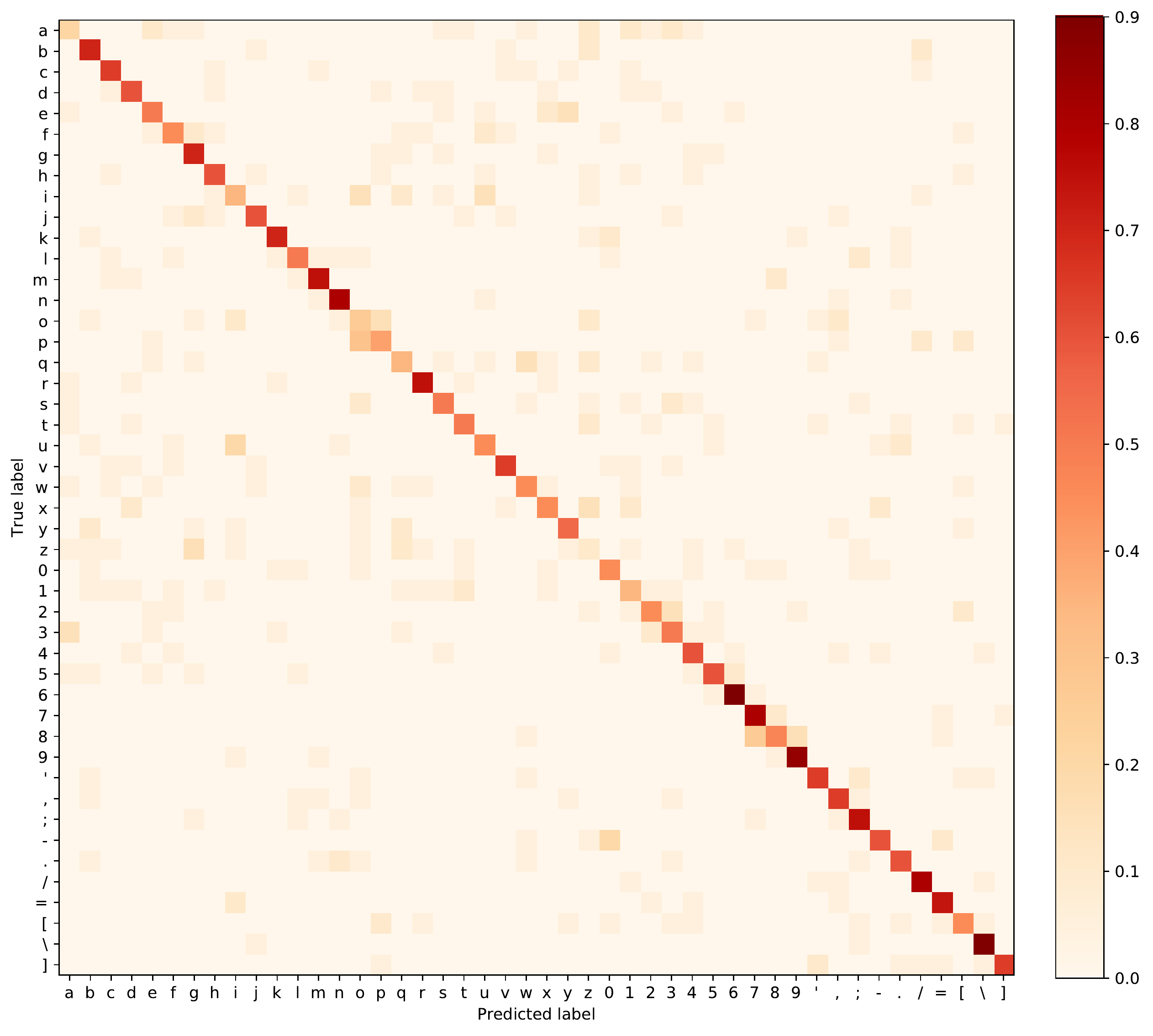}
\caption{\changed{Normalized confusion matrix example from Cross-Validation -- Logitech HPTT Model.}}
\label{fig:cross_v}
\end{figure}

\begin{figure}
	\begin{minipage}{.45\linewidth}
		\includegraphics[height=4cm,width=\linewidth]{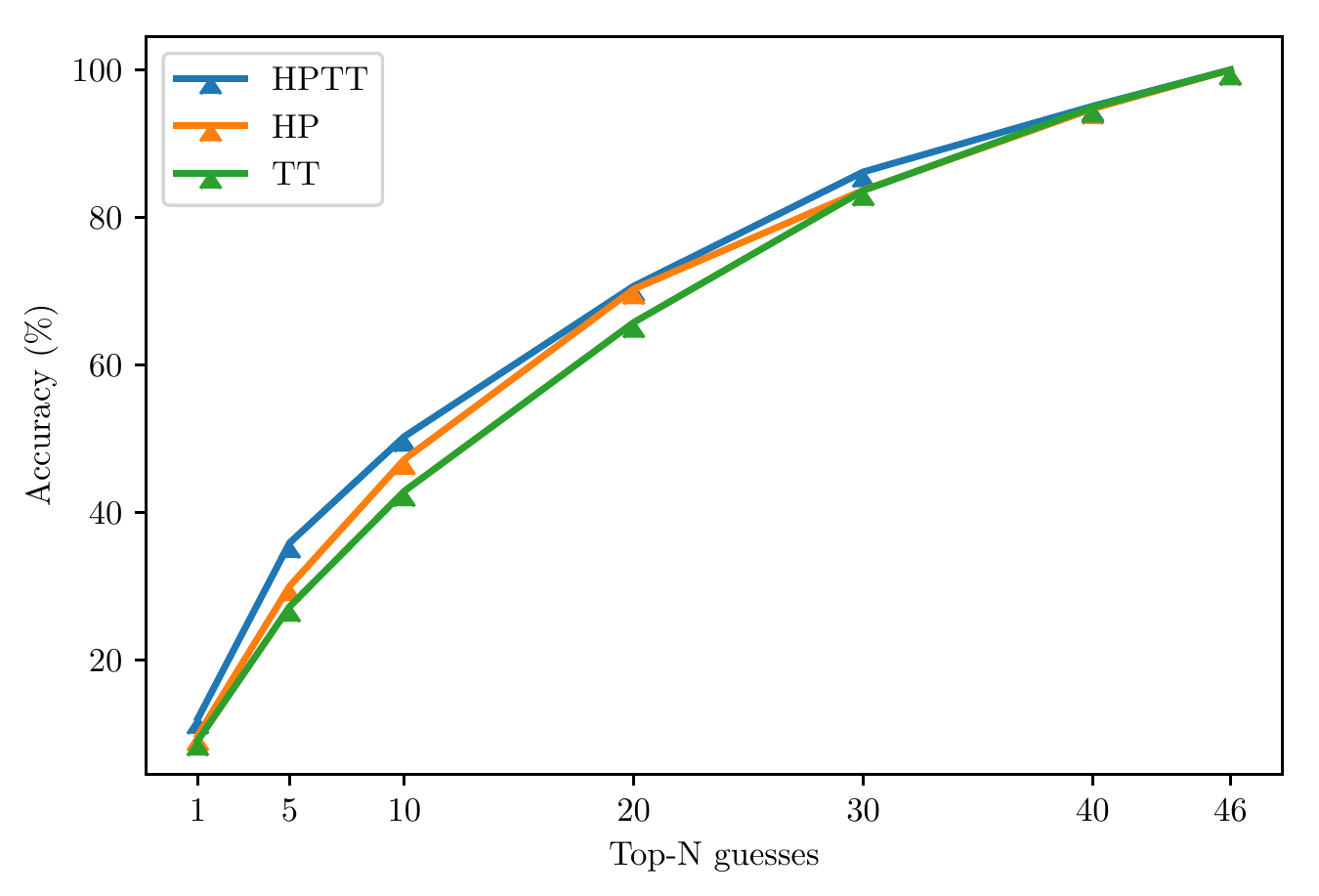}
		\caption{Top-N accuracies of our models: HP -- Hunt-and-Peck, TT -- Touch-Typing, HPTT -- both HP and TT.}
		\label{fig:topn}
	\end{minipage}
	\begin{minipage}{.45\linewidth}
			\includegraphics[height=4cm,width=\linewidth]{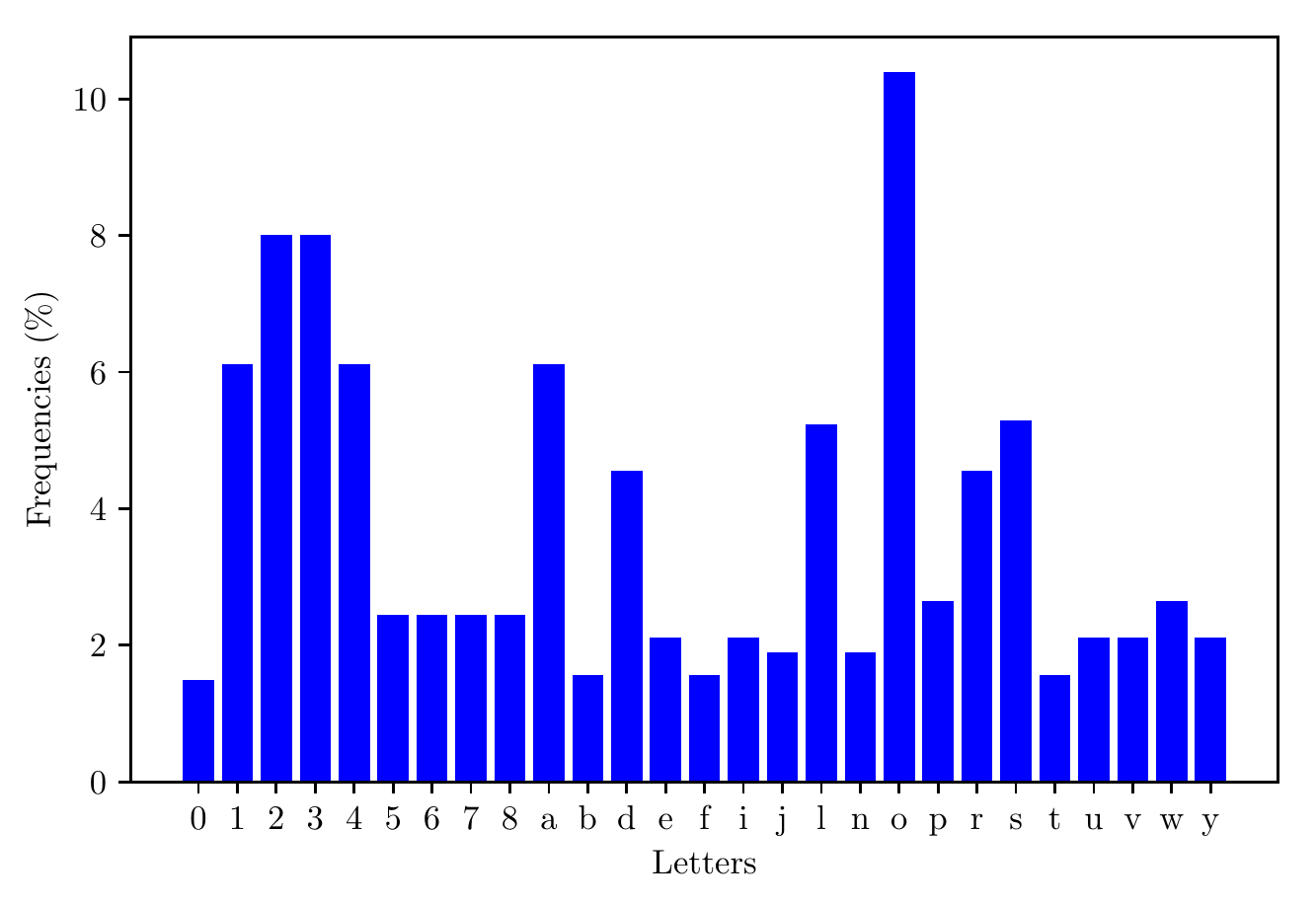}
		\caption{Aggregate letter frequencies of all passwords entered by our subjects.}
		\label{fig:freq}
	\end{minipage}
\end{figure}

\changed{According to results obtained in~\cite{compagno2017don}, we use a Logistic Regression classifier ($C = 1.0$, $penalty=`l2'$,$max\_iter=100$) to perform key classification, which outperformed Linear Discriminant Analysis, Support Vector Machines, Random Forest (RF), and k-Nearest Neighbors.}
%
Our experiments include 3 popular commodity plastic keyboards: Dell SK-8115, Logitech YUM76A, and 
AZiO Prism KB507. We generate HP, TT and HPTT models for each keyboard and use them to obtain 
keystroke guesses. 5-Fold Cross-validation scores for each model are shown in Table~\ref{tab:kb-cv}. Overall, Hunt-and-Peck models achieve higher success rates compared to Touch Typing models. 
\changed{A normalized confusion matrix example from the Cross-validation is presented in Figure~\ref{fig:cross_v}. Among the most misclassified keys, we find ``a'' is often confused with near keys such as ``z'' or ``s'', or the ``o'' and ``p'' keys. Symbols appear to have the higher prediction rate.}
A comparison of 
these models' accuracies over $1,400$ characters entered by study subjects is in Figure~\ref{fig:topn}. 
Letter frequencies for all passwords are shown in Figure~\ref{fig:freq}.  Since HPTT models achieve higher 
classification rates as shown in Figure~\ref{fig:topn}, we use the HPTT models (for the corresponding keyboards) 
in our evaluation.

\begin{table}[]
	\centering
	\small
	\caption{Cross-validation scores of HP, TT and HPTT models for each keyboard used in our experiments.}
	\label{tab:kb-cv}
	
\begin{tabular}{|c|c|c|c|}
\hline
\multicolumn{1}{|l|}{} & \multicolumn{1}{l|}{AZiO Prism KB507} & \multicolumn{1}{l|}{Dell SK-8115} & \multicolumn{1}{l|}{Logitech YUM76A} \\ \hline
HP                     & 71.9\%                                & 62.1\%                            & 61.7\%                               \\ \hline
TT                     & 49.8\%                                & 46.9\%                            & 50.7\%                               \\ \hline
HPTT                   & 61.5\%                                & 57.2\%                            & 58.7\%                               \\ \hline
\end{tabular}
\end{table}

\section{Combining Side-channels for Password Recovery}\label{sec:rec}
Various side-channels yield information on an individual key or a combination thereof. For example, approaches 
relying on keystroke characteristics~\cite{asonov2004keyboard,zhuang2009keyboard,compagno2017don} produce a 
list of candidate keys for a given keystroke. Meanwhile, others, such as those relying on inter-keystroke timings 
\cite{song2001timing,foo2010timing}, use timing of two adjacent keystrokes to produce candidate key-pairs that 
match timing statistics.

In this section, we describe how these sources of information can be combined using a graph-based 
mechanism, a simple version of which was proposed by~\cite{faezioligo} for a single side-channel. 
This mechanism creates the basis of 
\acutherm, discussed in detail in Section~\ref{sec:combine}.

Our graph-based side-channel combining method has the following properties:
\begin{compactitem}
	\item The graph includes $start$ and $end$ nodes. A candidate character at position $i$ in the password 
	is represented as a node at Layer $i$ in the graph.
	\item An edge between nodes $a$ and $b$ at layers $k$ and $k+1$, respectively, exists if  
	the digram $ab$ is viable at positions $k$ and $k+1$ of the password.
	\item $start$ and $end$ nodes have edges to/from Layer 1 and Layer $n$ nodes, respectively, 
	where $n$ is the password 
	length.
	\item The weight of an edge is the amount of information obtained from the two side-channels. It can be expressed as a 
	probability, confidence metric or any other numerical value.
\end{compactitem}
This structure allows easy change of graph properties, yielding efficient combination of various sources of information. 
For example, if the probability of $b$ occurring at position $(k+1)$ in the password is $0.7$, the weight of the edges 
from any node in Layer $k$ to $b$ at Layer $k+1$ is $0.7$. Furthermore, edge weights can be increased according to 
the likelihood of a key-pair (digram) at a given index.

\begin{figure}[ht!]
\centering
 \resizebox{.5\textwidth}{4cm}{
	\begin{tikzpicture}
	\tikzstyle{state}=[shape=circle,thick,draw,minimum size=1cm]
	\node[state] (v1) at (-3,0.5) {$start$};
	\node[state] (v2) at (-1,2.5) {$a$};
	\node[state] (v3) at (-1,0.5) {$b$};
	\node[state] (v4) at (-1,-1.5) {$c$};
	\node[below of = v4] {Layer 1};
	
	\node[state] (v5) at (1,2.5) {$a$};
	\node[state] (v6) at (1,0.5) {$b$};
	\node[state] (v7) at (1,-1.5) {$c$};
	\node[below of = v7] {Layer 2};
	
	\node[state] (v8) at (3,2.5) {$a$};
	\node[state] (v9) at (3,0.5) {$b$};
	\node[state] (v10) at (3,-1.5) {$c$};
	\node[below of = v10] {Layer 3};
	
	\node[state] (v11) at (5,0.5) {$end$};
	
	\draw[->, ultra thick]  (v1) edge node[above,left=0.1cm]{\textbf{0.7}} (v2);
	\draw[->]  (v1) edge (v3);
	\draw[->]  (v1) edge (v4);
	\draw[->]  (v2) edge (v5);
	\draw[->, ultra thick]  (v2) edge node[below=0.2cm,left=0.1cm]{\textbf{0.5 + 0.8}} (v6);
	\draw[->]  (v2) edge (v7);
	\draw[->]  (v3) edge (v5);
	\draw[->]  (v3) edge (v6);
	\draw[->]  (v4) edge (v7);
	\draw[->]  (v4) edge (v5);
	\draw[->]  (v4) edge (v6);
	\draw[->]  (v4) edge (v7);
	\draw[->]  (v5) edge (v8);
	\draw[->]  (v5) edge (v9);
	\draw[->]  (v6) edge (v10);
	\draw[->]  (v5) edge (v10);
	\draw[->]  (v3) edge (v7);
	\draw[->, ultra thick]  (v6) edge node[above=0.2cm,left=0.01cm]{\textbf{0.8 + 0.7}} (v8);
	\draw[->]  (v6) edge (v9);
	\draw[->]  (v7) edge (v8);
	\draw[->]  (v7) edge (v9);
	\draw[->]  (v7) edge (v10);
	\draw[->]  (v8) edge (v11);
	\draw[->]  (v9) edge (v11);
	\draw[->]  (v10) edge (v11);
	\end{tikzpicture}   
 	}
	\caption{A Directed Acyclic Graph of a 3-character password that consists of keys chosen from the set $\{a, b, c\}$. 
	The path for the best guess $aba$ is shown in bold. In the sums, the first addend is the probability to be in the node, 
	while the second is the likelihood to go in that specific next node.}
	\label{fig:dag}
\end{figure}
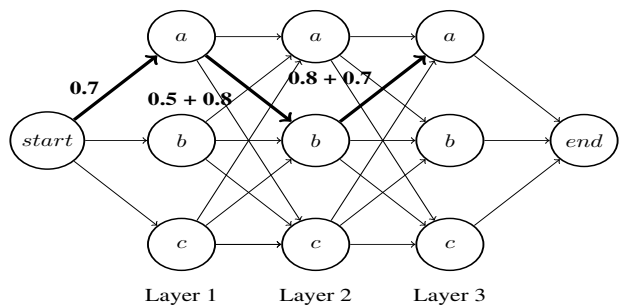

To combine the information from various side-channels, edge weights are changed to reflect the likelihood of the character or a 
digram occurring at a given position in the password. For example, to include the timing side-channel, statistical likelihoods of 
each digram at a given index can be used to increase or decrease the edge weight. An example graph for a 3-character 
password that consists of keys chosen from $\{a, b, c\}$ is shown in Figure~\ref{fig:dag}. An example probability distribution of 
keys and likelihoods of each digram at given positions is shown in Table~\ref{tab:prob}. 

This graph-based approach also allows us to adapt to various changes, including:
\begin{compactitem}
	\item \textbf{Change in possible characters at a given position: } This is handled by changing the nodes in the 
	respective layer and updating the edges.
	\item \textbf{Password length change:} More layers can be added or removed depending on password length.
	\item \textbf{Specific rules:} For example, if the timing signature does not coincide with a repeated keys (e.g., ``aa''),
	the edge between these two nodes can be removed, resulting in space reduction.
\end{compactitem}
After graph generation, most-likely passwords can be found efficiently using a k-longest path algorithm. In particular, 
the well-known Eppstein's algorithm~\cite{eppstein1998finding} has complexity of $\mathcal{O}(m + n\log n + k)$ 
where $m$ is the number of edges, $n$ is the number of nodes and $k$ is the number of paths.
Although this approach is suitable for combining multiple side-channels, thermal side-channel
reduces the password search space enough, such that the likelihood score for each password can be 
directly calculated (without the need to build a graph). In the next section, we describe this approach.

\begin{table}[ht!]
	\centering
	\caption{Example estimated probabilities of each character and likelihoods of key-pairs for each index.}
	\small
	\label{tab:prob}
	\begin{tabular}{|c|P{1cm}|P{1cm}|P{1cm}|}
		\hline
		\multirow{2}{*}{\textbf{Character}} & \multicolumn{3}{c|}{\textbf{Probability at Index}} \\ \cline{2-4} 
		& \textbf{1}      & \textbf{2}      & \textbf{3}     \\ \hline
		a                                   & 0.7             & 0.4             & 0.8            \\ \hline
		b                                   & 0.2             & 0.5             & 0.1            \\ \hline
		c                                   & 0.1             & 0.1             & 0.1            \\ \hline
	\end{tabular}
	\hspace*{1cm}
	\begin{tabular}{|c|P{1cm}|P{1cm}|P{1cm}|}
		\hline
		\multirow{2}{*}{\textbf{Index}} & \multicolumn{3}{c|}{\textbf{Likelihood at Index}} \\ \cline{2-4} 
		& \textbf{ab}      & \textbf{ba}      & \textbf{..}     \\ \hline
		1                                   & 0.8             & 0.1             & ..            \\ \hline
		2                                   & 0.2             & 0.7             & ..            \\ \hline
	\end{tabular}
\end{table}

\subsection{Combining Thermal \& Acoustic Side-Channels}\label{sec:combine}
We now describe the \acutherm attack and, as part of it, introduce a new password search space reduction mechanism, 
\textit{guided search} that mitigates the shortcomings of either side-channel.

\subsubsection{Guided Search}\label{sec:guided}
A password search space reduction method that assigns scores to each password in the password search
space. The search space is generated from two sources: (1) the set of keys obtained from a \attack attack, based on 
the thermal residue side-channel, and (2) password length leaked by the acoustic side-channel.
\changed{Since the \attack attack cannot infer the order of the keys or whether they are repeated, it is only possible to obtain the set of keys in the passwords using this side-channel.
Combined with the password key length information from the acoustic side-channel, the password search space is generated using the set of keys and their repetitions.
For instance, for a 3 character password and the password key set ${a, b}$, the password search space is comprised of ``aab'', ``aba'', ``abb'', ``baa'', ``bab'', and ``bba''.
}
For each password in the search space, password scores are calculated 
using predictions of the keyboard acoustics model for each key. Since the correct password -- assuming model predictions 
are accurate -- ranks closer to the top of the list, the attacker's likelihood of obtaining the correct password is 
increased. This space reduction is especially important for insider adversaries who do not want to trigger any mechanisms that monitor
excessive number of password attempts.

Our keyboard acoustics model essentially performs a classification task with 46 
classes, where each class is a keyboard key. 
It then returns a list of keys with confidence scores based on how likely the 
it is that each key matches the input sound. Overall, assuming 
that each classification task is independent, the confidence of a guess 
password being the correct one can be expressed as:
\begin{equation}
P(C) = \prod_{i = 1}^{|C|} P(c_i)
\end{equation}
where $C$ is a password guess, $|C|$ is password length, $c_i$ is a character 
and $P(c_i)$ is the probability obtained from the keyboard acoustics model. If the model's confidence for a key is high 
though not accurate, then it is likely that search space reduction might decrease drastically. To prevent this, we also 
include assigning scores based on linearly decreasing values. The option of 
using sum is introduced to better combine 
side-channels that might not be based on probabilities.

An instantiating of \textit{guided search} is shown in Algorithm~\ref{alg:char}. This password ranking method uses probabilities as 
character scores and multiplies them to combine these scores. We also tested the following methods for assigning likelihood 
values to each character:
\begin{compactitem}
	\item \textbf{Probability-Based:} Each character in the guessed password is assigned a score representing the probability of 
	being the correct character for the corresponding position, as returned from the keystroke characteristics model.
	\item \textbf{Linearly Decreasing Values (LDV):} Each character is assigned points according to its position 
	in the keystroke characteristics model. For example, if $46$ keys are returned from the model and the key in the guessed 
	password appears in the 0-th index, it is assigned 46 points.
\end{compactitem}
We tested two methods for combining likelihood scores for each key:
\begin{compactitem}
	\item \textbf{Sum:} Individual key scores are summed up to yield a final score for the password.
	\item \textbf{Multiply:} Each key score is multiplied to produce a final score for the password.
\end{compactitem}

\begin{algorithm}
	\caption{Password Score Based on Keyboard Acoustics}
	\label{alg:char}
	\SetAlgoLined
	\SetKwInOut{Input}{Input}\SetKwInOut{Output}{Output}
	\Input{\textit{guess\_password} $\gets$ A password guess}
	\Input{\textit{key\_predictions} $\gets$ A list of key predictions and probability pair from keyboard acoustics model (sorted from best to worst)}
	\Input{\textit{number\_of\_pred} $\gets$ Number of predictions from model}
	\Input{\textit{password\_length} $\gets$ Length of the guess password}
	\Output{A score based on Keyboard Acoustics}
	score = 1	\\
	\ForEach{$i = 1$ to password\_length}{
		\ForEach{$j = 1$ to number\_of\_pred}{
			\uIf{guess\_password[i] == prediction[j].key}{
				\tcc{Calculate probability of \textit{guess\_password} being the correct password.}
				score = score * prediction[j].prob\\
				\textbf{break}
			}	
		}	
	}
	return score
\end{algorithm}

\section{Results}
\label{sec:res}
In this section, we present the results of \system and \acutherma experiments.

\subsection{\attacka}
We consider the results of Stage Two analysis of thermal images obtained in Stage One. 
We divide them into two categories: 
\begin{compactitem}
\item  Hunt-and-Peck Typists --- `those who {\bf do not} rest their fingertips on, or hover their fingers just over, 
the home-row of keys (i.e. ``ASDF'' on the left hand, and ``JKL;'' on the right hand.). 
\item Touch Typists -- those whose fingertips routinely hover over, or lightly touch, the home-row. 
\end{compactitem}
The distribution among our 30 Stage One subjects was: 18 Hunt-and-Peck, and 12 Touch, typists.

%
%
As it turns out, study results indicate that the category of the typist is the most influential 
factor for the quality thermal imaging data. For each category, 
we separately analyze ``secure'' and ``insecure'' passwords types. Since we did not observe a significant statistical difference between results of different keyboards, results include all keyboards. 

\begin{figure}[!h]
\centering
\includegraphics[height = 1.2in, width=0.75\columnwidth]{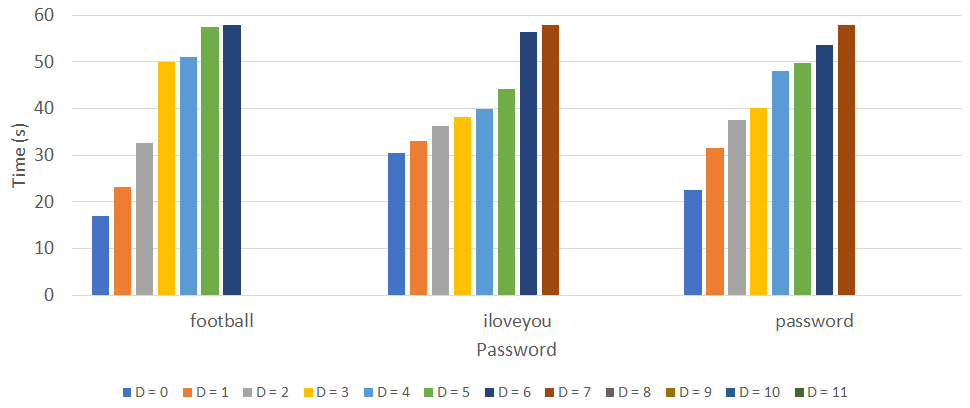} 
\caption{\small {Stage 2 Subject Performance: Alphabetical ``Insecure'' Passwords, all Typists.}}
\label{fig:abcsec}
\end{figure}

\begin{figure}[!h]
\centering
\includegraphics[height = 1.2in, width=0.75\columnwidth]{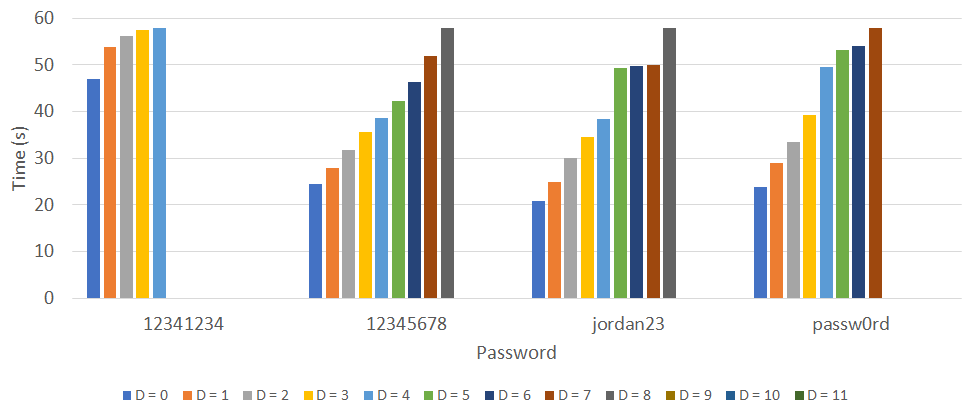} 
\caption{\small {Stage 2 Subject Performance: Alphanumeric ``Insecure'' Passwords, all Typists.}}
\label{fig:alphasec}
\end{figure}

\begin{figure}[!h]
\centering
\includegraphics[height = 1.2in, width=0.75\columnwidth]{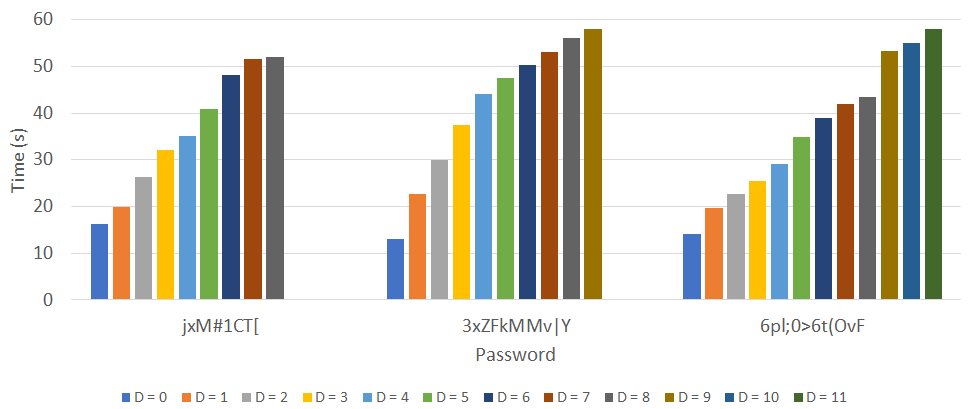} 
\caption{\small {Stage 2 Subject Performance: ``Secure'' Passwords, all Typists.}}
\label{fig:allsec}
\end{figure}

For full context, aggregate results (identification rates) from the entire subject population are shown in 
Figures~\ref{fig:abcsec},~\ref{fig:alphasec} and~\ref{fig:allsec}; they correspond to stage 2 subjects' analysis of 
``insecure'' and ``secure'' passwords, respectively. For clarity's sake, ``insecure'' passwords are split into two 
subcategories: alphabetical and alphanumeric. The former contains ``insecure'' passwords that consist only of 
English-language letters, while the latter contains  ``insecure'' passwords that 
also include numbers.  In each graph, ``D = 0'' refers to average latest time when stage 2 subjects could correctly identify 
every keystroke  of the entered password, while ``D = 1'' denotes average latest time when subjects could identify all-but-one 
keystroke; ``D = 2'' denotes the average latest time when subjects could identify all-but-two keystrokes, and so on. 
The distance "D" is calculated as:
\begin{equation}
D = \vert {(K \cup P) \setminus (K \cap P)} \vert
\end{equation}
where $P$ is the set of pressed keys identified by Stage 2 subjects and $K$ is the set of keys in the actual password. 
Note that keys missed and misidentified as pressed are considered in this distance calculation.

\subsubsection{Hunt-and-Peck Typists}
\label{subsec:handp}
The analysis of Hunt-and-Peck typists was straightforward. Because such typists do not rest their fingertips 
on (or hover right above) the keyboard home-row, it is readily apparent that each bright spot on the thermal image 
corresponds to a key-press. However, as discussed below, we encountered some challenges with ``secure'' passwords.

\begin{figure}[!h]
\centering
\includegraphics[height = 1.2in, width=0.75\columnwidth]{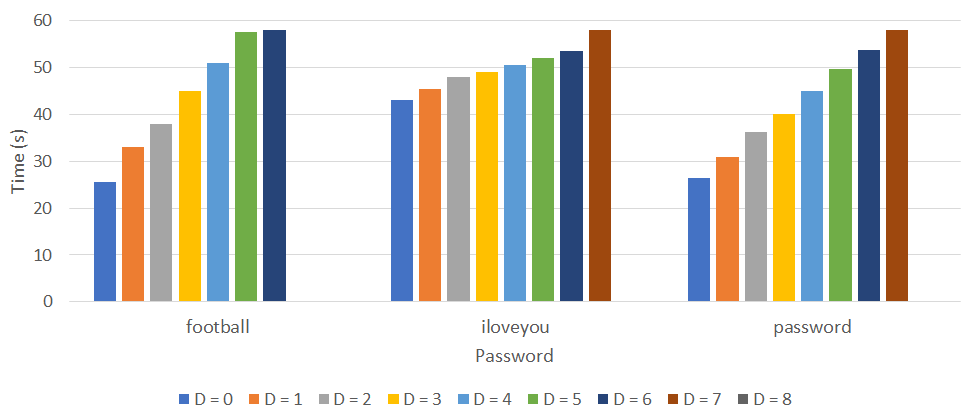} 
\caption{\small {Stage 2 Subject Performance: Alphabetical ``Insecure'' Passwords, Hunt-and-Peck Typists.}}
\label{fig:handpabc}
\end{figure}

\begin{figure}[!h]
\centering
\includegraphics[height = 1.2in, width=0.75\columnwidth]{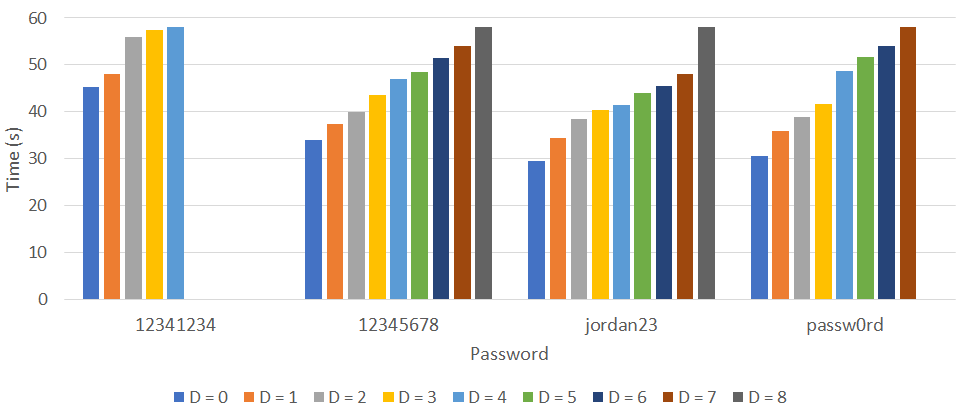} 
\caption{\small {Stage 2 Subject Performance: Alphanumerical ``Insecure'' Passwords, Hunt-and-Peck Typists.}}
\label{fig:handpnum}
\end{figure}

\paragraph{Insecure Passwords}
As Figure~\ref{fig:handpabc} and~\ref{fig:handpnum} show, analysis of Hunt-and-Peck typists entering ``insecure'' 
passwords is trivial. 
In fact, in the best-case of ``12341234'' subjects could correctly recall every keystroke, on average, $45.25$ seconds after 
entry. Even the weakest result, ``football'' was fully recoverable $25.5$ seconds later, on average. This is in line with 
conventional thought. Hunt-and-Peck typists typically only use their forefingers to type. Because of this, they make contact 
with a larger finger over a large surface area. Also, since Hunt-and-Peck typists are generally less skilled, they take 
longer for each keystroke, resulting in longer contact time. These two factors combined yield high-quality thermal residue 
for \attacka .


\begin{figure}[h!]
\centering
\includegraphics[height = 1.2in, width=0.75\columnwidth]{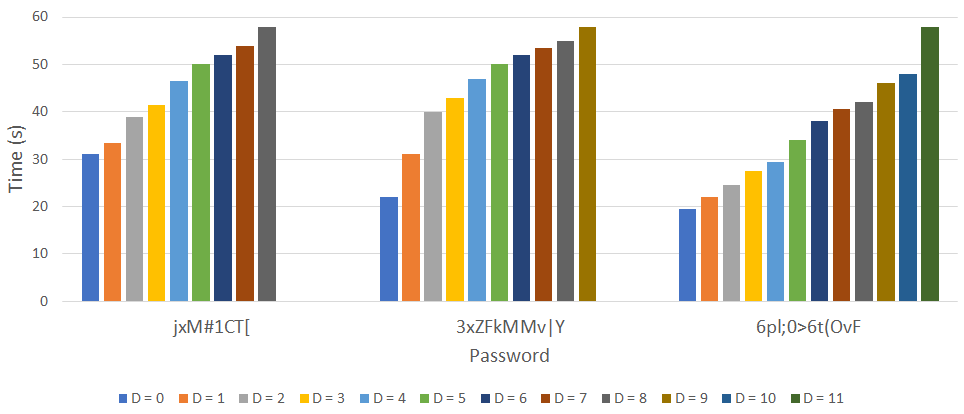} 
\caption{\small {Stage 2 Subject Performance: ``Secure'' Passwords, Hunt-and-Peck Typists.}}
\label{fig:handpsec}
\end{figure}

\paragraph{Secure Passwords}
``Secure'' passwords are more challenging to analyze. As shown in Figure~\ref{fig:handpsec} full recall was possible, 
on average, 
up to $31$ seconds after recording started, in the best case, and $19.5$ seconds, in the worst case. Performance of stage 2 
subjects was uniform in terms of password length: the shortest password was the easiest to analyze correctly. 
Anecdotally, this is not surprising. It was quite common for Hunt-and-Peck typists to look back and forth between 
the characters
of a relatively complex ``secure'' passwords, and their keyboards. This resulted in longer completion times, 
which left longer time for keycaps to cool off before recording began.

\subsubsection{Touch Typists}
\label{subsec:touch}
Analyzing data from Touch typists was a challenge for stage 2 subjects. Since a typical Touch typist's fingers are 
constantly in contact with (or in very close proximity of) the home-row of the keyboard, there are two incidental 
sources of thermal noise. First, there is thermal residue on the 2 groups of 4 home-row keys: ``asdf'' and ``jkl;'' 
which results from the typist's fingertips. However, whenever typist's fingers rest on the keyboard for a long time,
additional observed effects occur outside (though near) the home-row, on the following keys:

\fbox{{\centering\noindent\bf
{\tt "qwertgvcxz"}, {\tt "][poiuhnm,./"}
}}
\\
Even though this secondary thermal residue was not as drastic as that on the home-row, it had a more pronounced 
effect on stage 2 subjects. In many cases, a subject was uncertain whether a key was lit on the 
thermal image because it was actually pressed, or because it was simply close to the home-row. 
This uncertainty in turn led to mis-classification of some keys as unpressed. Also, mis-classification of 
home-row keys as pressed keys was not counted in the distance. We justify this choice in Section~\ref{sec:dis}.

\begin{figure}[h!]
\centering
\includegraphics[height = 1.2in, width=0.75\columnwidth]{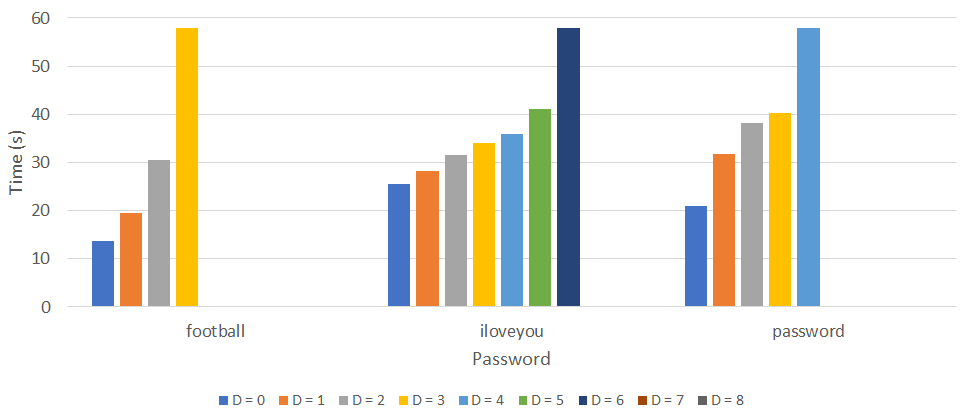} 
\caption{\small{Stage 2 Subject Performance: Alphabetical ``Insecure'' Passwords, Touch Typists.}}
\label{fig:abctouch}
\end{figure}

\begin{figure}[h!]
\centering
\includegraphics[height = 1.2in, width=0.75\columnwidth]{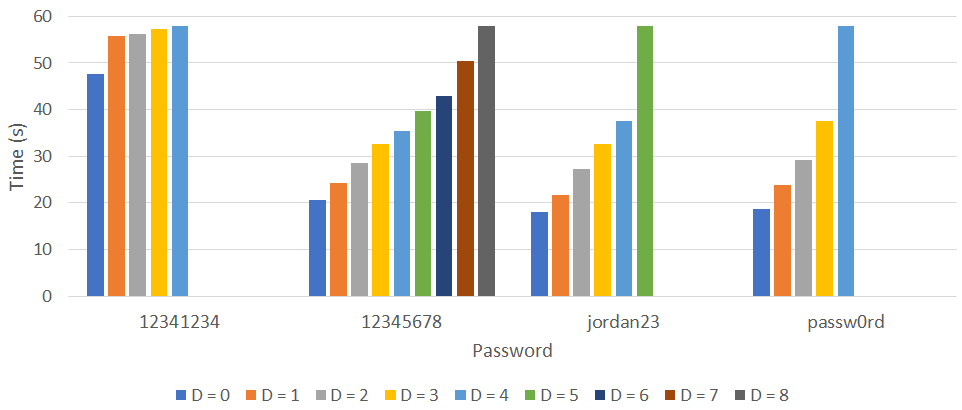} 
\caption{\small{Stage 2 Subject Performance: Alphanumeric ``Insecure'' Passwords, Touch Typists.}}
\label{fig:alphatouch}
\end{figure}

\paragraph{Insecure Passwords}
While more difficult than analysis of ``insecure'' password for Hunt-and-Peck typists, stage 2 subjects have
moderate success analyzing Touch typists entering ``insecure'' passwords. As Figures~\ref{fig:abctouch} and 
\ref{fig:alphatouch} 
show, the best average time for full recall was for password: ``12341234'' at $47.6$ seconds, and the worst 
was for ``jordan23'', at $17.8$\ seconds. This follows the notion that stage 2 subjects were hesitant to classify 
home-row-adjacent key-presses, e.g., ``o'', ``r'' and ``n'' in ``jordan23''. Furthermore, this supports the notion that a 
simple, repeated password such as ``12341234'' leaves ideal thermal residue. Since each key is repeated, 
it is analogous to each key being pressed once for twice as long. This results in twice as much thermal energy 
being transferred from the fingertip to the keycap.


\begin{figure}[h]
\centering
\includegraphics[height = 1.2in, width=0.75\columnwidth]{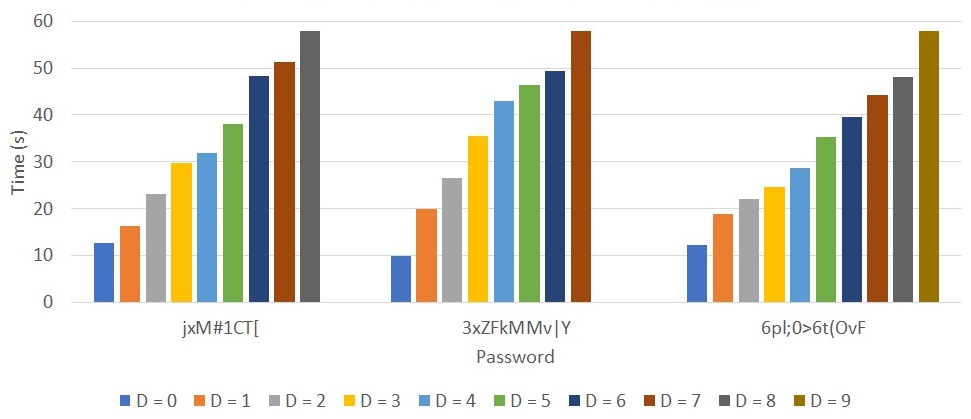} 
\caption{\small{Stage 2 Subject Performance: ``Secure'' Passwords, Touch Typists.}}
\label{fig:touchsec}
\end{figure}

\paragraph{Secure Passwords}
Touch typists entering ``secure'' passwords were the most difficult for the stage 2 subjects to analyze. 
As shown in Figure~\ref{fig:touchsec}, full recall was only possible, on average, within the first $14.33--18.5$ 
seconds. Surprisingly, the password with the smallest window for full recall was ``jxM\#1CT[''.  We believe 
that many stage 2 subjects were hesitant to classify home-row-adjacent keys in this password as keystrokes
(as opposed to thermal noise). This might explain why the window for full recall is so small. As with all 
other cases, the time window between full recall at $d=0$ and a single mis-identification $d=1$ was much 
greater than any other window between $d=n$ and $d=n+1$, which is consistent with Newton's Law of Cooling.

\subsubsection{Outlier: Acrylic Nails}
There was a single Stage 1 subject that had long acrylic fingernails. Instead of typing with fingertips, this person
tapped the keys with nail-tips. Since these do not have nearly as much surface area as fingertips, 
and false nails do not have any blood vessels to regulate their temperature, this subject left almost no thermal residue. 
In fact, not a single key-press could be correctly identified in any of the $40$ password entry trials. Consequently, 
this subject is not included in either  Touch or Hunt-and-Peck typist populations. However, as a side curiosity, we
note that, although it may be a rare occurrence, any user with long acrylic fingernails is virtually immune to \attacka.

\subsection{\acutherma}

\changed{Through \attacka, an attacker gains information on which keys have been pressed, but nothing about their orders or the password length. Thus, to guess what the password actually is, the attacker should try all the possible combinations of the keys resulting from the thermal image with all the possible lengths, i.e., the password search space is infinite. By leaking the password length through acoustic emanations, we can reduce the search space drastically. Once the search space is defined, the ordering of key-presses is the second factor of space reduction. Thus, to evaluate our approach of combining thermal and acoustic side-channels, we create the password search space for each password based on the length and the key-set (i.e., a finite number), and we calculate the space reduction by using the acoustic information to reconstruct the key ordering. In other words, we evaluate how easier it is to guess the password by exploiting thermal and acoustic information rather than only thermal plus password length information.}To rank candidate  
passwords, we assign scores to each, as described in Section~\ref{sec:guided}. 
We now show the results of password 
search space reduction, which is defined as:
\begin{equation}
1 - \frac{l}{|P|}
\end{equation} 
where $l$ is the rank of the correct password in the list sorted (based on scores) and $P$ is the password search space size. 
Reductions are given over the password search space reduced using the \attacka.

Columns labeled ``Keystroke'' in Tables~\ref{tab:sum_red} and~\ref{tab:mul_red} show the password search space reduction 
percentages of using sum and multiplication as the combination of scores for each key. Multiplication and sum combination 
methods perform equally well, except for multiplication of LDV. Use of probabilities compared to LDV results in higher 
space reduction, though the difference is not big enough to draw conclusions.

\begin{table}[ht!]
	\renewcommand\arraystretch{1.5}
	\centering
	\small
	{\caption{Password Space Reduction for Sum of LDV and Probabilities with Same Key Timings (increasing guess password score if password contains same key pressed twice and timings match).}
		\label{tab:sum_red}}
	\begin{tabular}{l|c|c|}
		
		\hline
		\multicolumn{1}{|l|}{\textbf{Model}}       & \textbf{Keystroke} & \textbf{Keystroke + Same Key Timings} \\ \hline
		\multicolumn{1}{|l|}{Sum of LDV}           & 81.8\%             & 83.6\%                                \\ \hline
		\multicolumn{1}{|l|}{Sum of Probabilities} & 83.2\%             & 84.3\%                                \\ \hline
	\end{tabular}
\end{table}

\begin{table}[ht!]
	\renewcommand\arraystretch{1.5}
	\centering
	\small
	\caption{Password Space Reduction for Multiplication of LDV and Probabilities with Same Key Timings.}
	\label{tab:mul_red}
	\begin{tabular}{l|c|}
		\hline
		\multicolumn{1}{|l|}{\textbf{Model}}                  & \textbf{Keystroke + Same Key Timings} \\ \hline
		\multicolumn{1}{|l|}{Mult. of LDV}           & 79.6\%                                \\ \hline
		\multicolumn{1}{|l|}{Mult. of Probabilities} & 83.7\%                                \\ \hline
	\end{tabular}
\end{table}

\begin{figure}[!ht]
	\includegraphics[width=\linewidth]{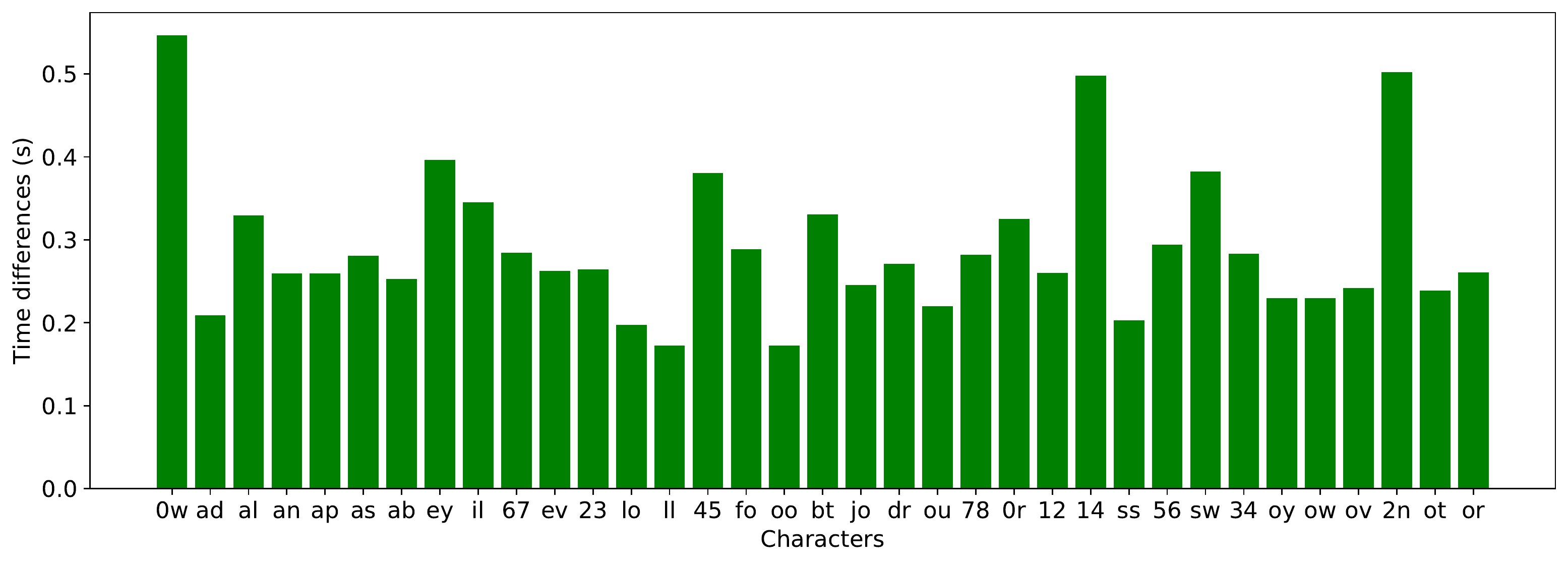}
	\caption{Interkeystroke timings (digraphs)} \label{fig:digraph}
\end{figure}

Although timings directly can not be used in score calculation since we do not have all key-pairs in our dataset, the observation 
that same-key presses are often typed fast (around 0.15 seconds See Figure~\ref{fig:digraph}) can be used to increase the score 
of a guessed password. 
This is done by increasing the score by 46 and 1 in for LDV and probabilities used as key scores, respectively. The column labeled
``Keystroke + Same Key Timings'' refers to the additional use of this observation. This approach results in about a 
1.4\% increase, on average, in reduced password space.

\begin{figure}[ht!]
	\includegraphics[width=\linewidth]{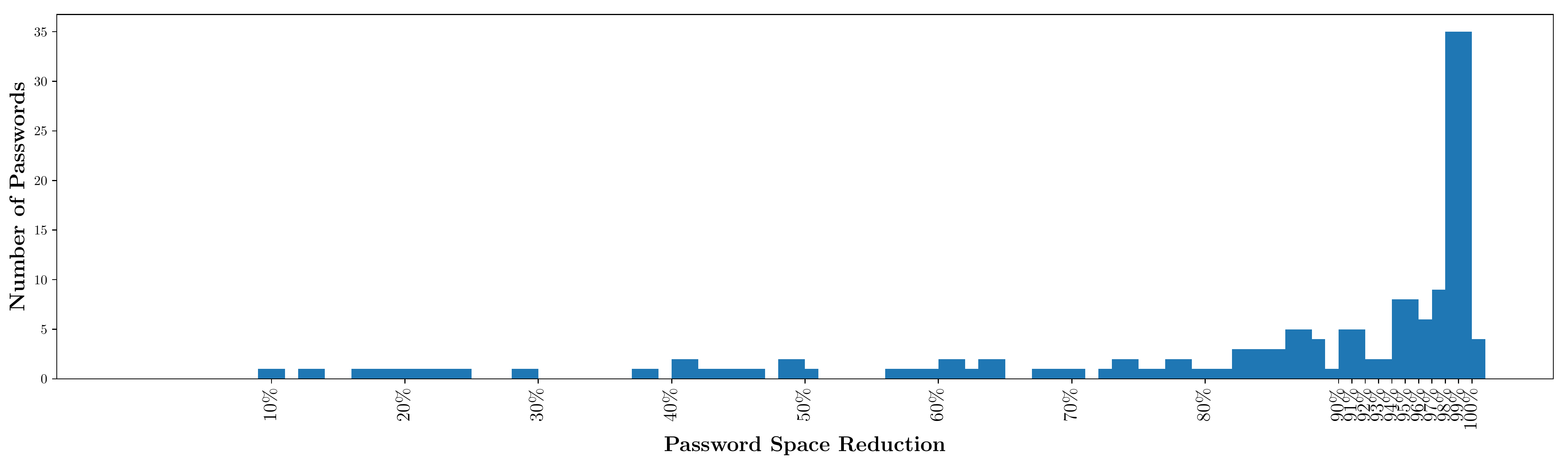}
	\caption{{Password Space Reduction for 130 passwords. 4 
			passwords were at the top of our guessed list. 
			39 and 67 passwords were found in Top 1\% and Top 5\%, respectively.}}
	\label{fig:red}
\end{figure}

\begin{table}[ht!]
	\centering	
	\small
	\caption{Password frequencies used in evaluation.}
	\label{tab:passwordfreq}
	\begin{tabular}{|c|c|c|P{1cm}|c|c|c|P{1.1cm}|}
		\hline
		\textbf{Password} &	password & 12345678 & football & iloveyou & 12341234 & passw0rd & jordan23 \\ \hline
		\textbf{\# of Entry} & 10 & 30 & 15 & 19 & 24 & 12 &21 \\ \hline
	\end{tabular}
\end{table}

We also give the individual password space reduction percentages for 130 passwords in Figure~\ref{fig:red}. Out of 130 
passwords entered by our 19 subjects, 4 correct passwords were at the top of our ranked password list and 
39, 48, 54 correct passwords were found in top 1\%, 2\% and 3\% of the password 
space, respectively. The distributions of these 
passwords are given in Table~\ref{tab:passwordfreq}. Per-subject entry 
varied between 1 and 5 for 6 subjects and for 
the other 13 subjects, the number of entered passwords ranged from 6 to 18 with 
an average of 9 and 
standard deviation of 3.3.

Space reductions for each keyboard are shown in Table~\ref{tab:red_per_kb}. While overall space reduction for 
AZiO keyboard is lower (77.3\%), other keyboards have a similar reduction 
($\approx$87\%). Though this may not completely coincide with 
cross-validation scores in Table~\ref{tab:kb-cv}, the lower reduction rate for 
the AZiO keyboard might be due to the model being confused with 
sounds that are frequent in our password set. From the cross-validation scores, we can see that touch-typing 
often results in lower key identification rates.

\begin{table}[ht!]
	\renewcommand\arraystretch{1.5}
	\small
	\centering
	\caption{{Space Reduction per Keyboard}}
	\label{tab:red_per_kb}
	
	
	\begin{tabular}{|c|c|c|}
		\hline
		\textbf{Logitech Y-UM76A} & \textbf{Dell SK-8115} & \textbf{AZiO Prism KB507} \\ \hline
		  86.9\% & 86.7\%  & 77.3\%  \\ \hline
	\end{tabular}

\end{table}


\subsubsection{Comparison With Similar Attacks}
\changed{Password recovery by exploiting acoustic emanations has been extensively studied in the literature~\cite{asonov2004keyboard}, but this side-channel was never combined with others before. Dictionaries often assist acoustic attacks in boosting performance~\cite{berger2006dictionary}; however, this is not suitable in the case of random text. \acutherm greatly overcomes this limitation by leveraging thermal emanations, which reveal the exact pressed keys. Compared to~\cite{compagno2017don}, in which the authors cope with randomness using an improved brute force approach, \acutherm is significantly more effective, since the set of keys to generate the randomness is well defined by the thermal image, rather than being the whole set of available keys. Moreover, the attack was evaluated on 19 participants using three different keyboards, with the good results highlighting more generalization than previous works~\cite{zhu2014context,10.1145/2789168.2790122}.

The most recent comparable work to ours, in terms of exploiting the thermal-residue side-channel, is~\cite{mowery2011heat}. 
This work focused on recovering PINs entered on a PIN pad.
In comparison to recovering passwords, recovering PINs is a relatively easy problem at first sight due to the known length of a PIN ($\approx4$ digits) and, in result, possibly shorter entry times.
Yet, the same work showed that discovering the order of key presses is hard. Perfect PIN recovery were only 10\% right-after entry and less by the time passed.
Our results confirm this for passwords as well.
However, by using an acoustic side-channel, we obtain the length of the password and predictions that aid in these two drawbacks of thermal-residue side-channels -- which have not been explored before.}

Sections~\ref{sec:rand} and~\ref{sec:nonrand} below discuss how to use these results to aid in password search space reduction 
for ``secure'' and ``insecure'' passwords, respectively.
\section{Discussion}
\label{sec:dis}
%
We now break down our observations from Section~\ref{sec:res} between  
two password classes, and among two categories of typists for \attacka. We also discuss how keyboard 
acoustics part of \acutherma performs on its own against ``secure'' and ``insecure'' passwords. 
We also discuss possible additional side-channels.

\subsection{\attacka}
\subsubsection{Results with ``Insecure'' Passwords}
Stage Two subjects were particularly adept at identifying passwords that are English words or phrases. Even though we 
could not reliably detect the exact sequence of pressed keys, ordering can be found indirectly by mapping the set of 
pressed keys to words (essentially, solving an anagram puzzle). Furthermore, a list of distances between 
detected keys (characters) and possible words, can be used to reconstruct full passwords from incomplete 
thermal residues.. 
Finally, the same list of distances can help determine when a key is pressed multiple times. These combinations 
highlight the threat posed by \attacka to already insecure passwords. 

\subsubsection{Results with ``Secure'' Passwords}
However, strong results from Stage 2 subjects' identification of English-language words does not extend to secure, 
randomly-selected passwords. First, inability to reliably determine the order of pressed keys can not be mitigated 
by leveraging the underlying linguistic structure. Moreover, it is unclear whether a given set of emanations represents the 
whole password, or if some information was lost. Finally, it is impossible to tell if a key was pressed multiple times. 
However, even with these shortcomings, our subjects managed to greatly reduce the password search space  
from $72^n$ to $72^{n-m}*m!$ where $n$ is the total number of characters in the password, and $m$ is the number 
of identified key-presses.  This represents a reduction in search space by a factor of $10^{10}$ for an $8$-character 
password where the individual keys have been identified. Techniques to further reduce the space of candidate 
passwords are discussed in the following section.

\subsubsection{Results with Hunt-and-Peck Typists}
\begin{figure}[th!]
	\centering
	\begin{minipage}{.45\linewidth}
		\includegraphics[height=1.0in,width=0.95\columnwidth]{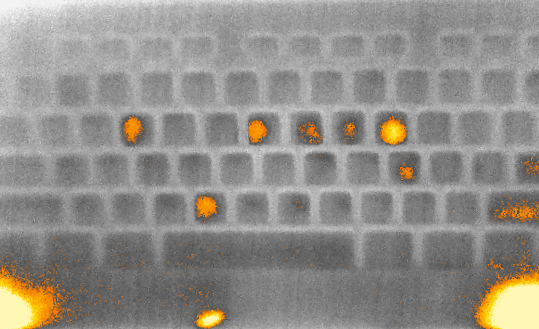} 
		\caption{{\small Password ``iloveyou'' entered by a Hunt-and-Peck typist.}}
		\label{fig:HandP}
	\end{minipage}
	\begin{minipage}{.45\linewidth}
		\includegraphics[height=1.0in,width=\columnwidth]{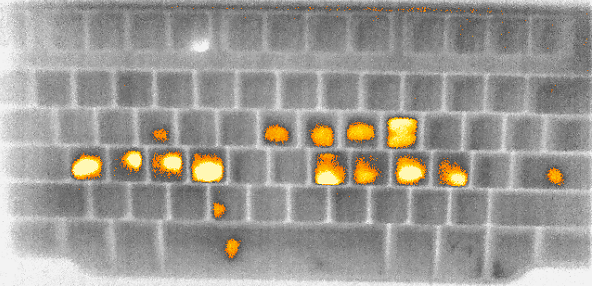} 
		\caption{{\small Password ``iloveyou'' Entered by a Touch Typist.}}
		\label{fig:home-row}
	\end{minipage}
\end{figure}
As described in Section~\ref{subsec:handp}, Hunt-and-Peck typists are particularly vulnerable to \attacka. 
This is not surprising, given that these less-skilled typists tend to type more slowly, and primarily use their index 
fingers, which usually have a greater fingertip surface area than ring or pinky fingers~\cite{peters2009diminutive}. 
This results in greater heat transfer, due to longer contact 
duration with a larger contact area. Also, as seen from Figure~\ref{fig:HandP}, Hunt-and-Peck typists do not touch
any keys that are not part of the password. Therefore, every observed key-press is part of the password.

\subsubsection{Results with Touch Typists}
For Touch typists, two factors confuse their thermal residues and make passwords harder to harvest. 
One is their habit to rest their hands on the home-row, which introduces potential false positives. as Figure~\ref{fig:home-row} 
shows. This is exacerbated by the possibility that any home-row key might actually be part of the password. Because of this, 
stage 2 subjects were not penalized for classifying the home-row keys as pressed; they were instructed to identify all 
keys that looked to them as having been pressed. 

Another issue is that Touch typists tend to use all fingers of both hands while typing. This causes two advantages over
their Hunt-and-Peck counterparts. First, they touch individual keys for a shorter time, thus transferring less 
heat to the key-cap. Second, they type much more quickly and also use their ring and pinky fingers. 
Fingertips of these smaller fingers tend to have $1/2$ of the surface area of larger index or middle fingers. 
Thus, they transfer half of the total heat energy due to conduction during a key-press~\cite{peters2009diminutive}. 
Such factors make Touch typists much more resistant to \attacka, particularly, 
at the level of our moderately sophisticated adversarial model.

\subsubsection{Ordering of Key-Presses}
%

\begin{figure}[th!]
	\centering
		\includegraphics[height=1.8in,width=\columnwidth]{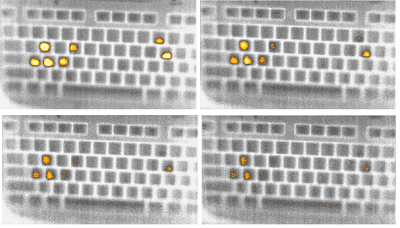} 
	\caption{{\small ``passw0rd'' thermal residue after 0 (top left), 15 (top right), 
	30 (bottom left), and 45 (bottom right) seconds after entry.}}
	\label{fig:timelapse}
\end{figure}

Unfortunately, inspection of thermal images by stage 2 subjects did not yield any reliable key-press ordering information. 
Newton's Law of Cooling might seem to indicate that any reduction in heat energy would occur uniformly across 
all pressed keys, resulting in exposure of ordering. However, this is not true in practice. One reason is due to by 
\emph{keystroke inconsistency} in the dynamics of Touch typists. Factors, such as the travel distance between keys 
and the particular finger used to press a key, result in small differences in the duration, and total surface area of, contact. 
Since each key-press is distinct, intensity of a given thermal residue does not correspond to its relative position in the 
target password. This holds even for Hunt-and-Peck typists, who tend to use only their index fingers. As evidenced 
by Figure~\ref{fig:timelapse}, Hunt-and-Peck typist does not necessarily press keys with uniform force or for a uniform duration. 
These inconsistencies make reliable ordering of key-presses infeasible in our analysis framework. 
However, as mentioned above, for insecure (language-based) passwords, dictionary tools can be used to infer the 
most likely key-press order.   

\subsection{\acutherma}

\subsubsection{``Secure'' Passwords}\label{sec:rand}
Random passwords correspond to keys selected randomly from a uniform distribution of keys. Since they lack structure 
(unlike dictionary words) that can be used to guide password search (e.g., using a dictionary), they require the whole 
password space to be explored, which can be quite large. For example, ``passw0rd'' with the knowledge of password length and 
a full set of keys: \{p, a, s, w, 0, r, d\}, has the search space of $141,120$ possible passwords. An adversary can  
guess this password after, on average, $70,560$ guesses.

Application of our results to ``secure'' passwords is straightforward since we do not rely on any dictionaries. For example, 
\textit{guided search} can be directly used for ``secure'' passwords to reduce the search space. If it is not possible to generate 
the whole password space, methods described in Section~\ref{sec:rec} can be used to match the graph structure with the 
information available about the password. The same section also discusses how most-likely passwords can be 
obtained using an efficient k-longest-path algorithm.

To apply methods discussed in Section~\ref{sec:rec} to touch typists, each layer in 
the created graph includes all hot keys on the thermal image classified as recently pressed. Edges are generated between 
layers and probabilities received from the keyboard acoustics model are assigned as the directed edge weight to the corresponding 
key in the correct layer. Then, any efficient k-longest-path algorithm can be used to obtain most-likely passwords.

\subsubsection{``Insecure'' Passwords}\label{sec:nonrand}
For ``insecure'' passwords such as the ones consisting of words, password dictionaries can be easily employed and previous work has shown that this approach is viable~\cite{berger2006dictionary, compagno2017don, narayanan2005fast}. We performed the following experiment to determine how successful an \acutherm attack is against an insecure password. This attack can even be used as a preprocessing step and be used to pick a substantially smaller set of passwords from a password dictionary to perform the attack described in Section~\ref{sec:combine}.

\subsubsection{Acoustic-only Side-channel Search Space Reduction}
\acutherm combines acoustic and thermal residue side-channels to reduce the password search space. In this paper, to obtain the combined password search space reduction, we apply the information collected from the acoustic side-channel to the thermal residue side-channel password search space. For an acoustic-side-channel-only version of this attack, we expect the password search space reduction to be similar (for instance 83.2\% for Sum of Probabilities without timings matching, see Table~\ref{tab:sum_red}) but with a much larger initial password search space. For instance, with a password length of 8 characters and 46 characters available, the final searching space of an acoustic-only side channel attack contains $46^8*(1-0.832) = 3,367,998,900,000$ passwords. 

Another approach for utilizing acoustic side-channel information is to only use the Top-N keys obtained from the classifiers. Although this approach does not guarantee that the password will be in the search space, it is mainly used to reduce the size of the password search space. With $N = 20$ which corresponds to 70\% probability of finding the correct key in our candidates for HP typists  (See Figure~\ref{fig:topn}), in a password with 8 characters, the final space is $20^8*(1-0.832) = 4,300,800,000$ passwords. Unfortunately, the probability of the correct password occurring in this space is only $ 0.70^8 =5.76\%$. The combination of both acoustic and thermal information is far more effective than using thermal or acoustic solo information. 

\subsubsection{\acutherm Dictionary Attack}
A password dictionary is typically a file where common passwords are listed in decreasing order of popularity. Since password length
can be inferred from the acoustic side-channel, and the pressed keys -- from the thermal image, we can use this information to 
efficiently search for the target password in the dictionary. First, we remove all passwords that do not have the same length as 
the target password. This leads to a significant password search space reduction. Second, we can calculate the key-set of each 
password in the dictionary, and compare it to the key-set extracted from the thermal image. We can define the distance between
two key-sets as the number of elements that do not appear in both sets (i.e., if the distance is zero, then the two sets are equal). To evaluate how well \acutherm works for ``insecure'' passwords, 
we measure the accuracy of finding the target password among Top-N guesses. 

As the password dictionary, we use the {\it phpbb.txt} database\footnote{This dictionary is available at: \url{https://wiki.skullsecurity.org/Passwords}},
containing $184,389$ passwords leaked from the {\it phpbb} website, sorted by 
popularity, in decreasing order. For each of $10,000$ 
most popular passwords in this dictionary, we compute the distance to every password (including the target password itself, with distance 
zero), and list them sorted by ascending distance. We then find the index of 
the password in that list, which 
gives us an idea on how many guesses, or Top-N guesses, are needed on average to find the correct password. We perform this 
experiment for both Hunt-and-peck Typists and Touch Typists.  

For a Hunt and Peck typist, the key-set extracted from a thermal image coincides with the key-set of the password, i.e., the distance is zero. 
However, for a Touch Typist this is not true, since home row keys (``a'',``s'',``d'',``f'',``j'',``k'',``l'',``;'') also appear in the thermal image. Thus, 
we add the home-row keys to the target password key-set before starting the password search. Note that the evaluation assumes that 
the target password is present in the dictionary.

\begin{figure}
	\centering
	\begin{minipage}{.55\linewidth}
	\centering
		\includegraphics[height=1.5in]{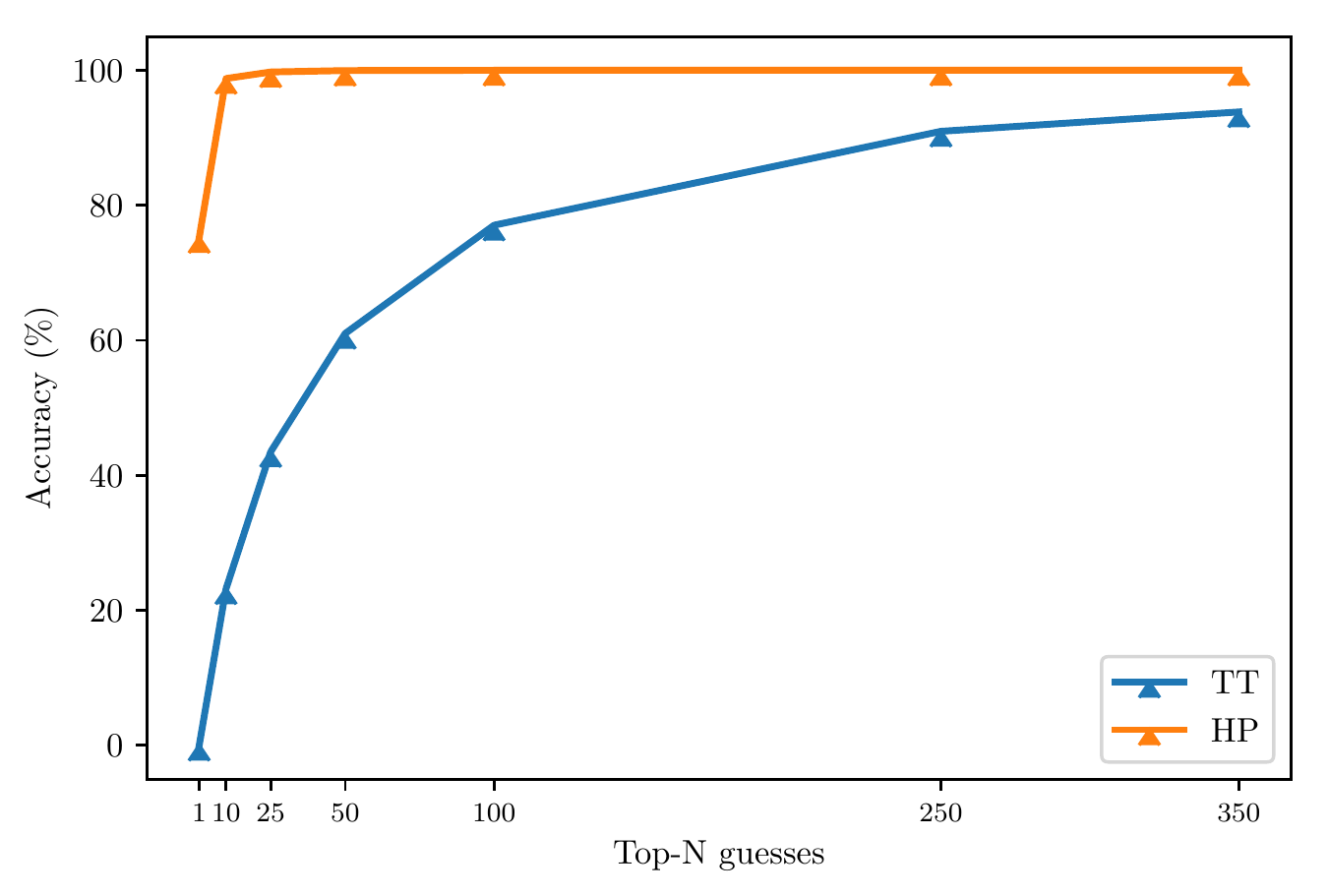}
		\caption{Top-N guess results for \acutherm Dictionary Attack against Hunt-and-peck and Touch Typists}
		\label{fig:top-N}
	\end{minipage}
	\hspace{0.5cm}
	\begin{minipage}{.4\linewidth}
		\centering
		\includegraphics[height=1.4in]{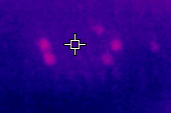} 
		\caption{\small{``football'', $10$ seconds after entry, captured by FLIR ONE.}}
		\label{fig:TG165}
	\end{minipage}
\end{figure}

Figure~\ref{fig:top-N} shows the accuracy of finding the target password with respect to different Top-N values. For Hunt-and-Peck typists, 
since the passwords are all found within distance 0, the initial accuracy is around 75\% for Top-1 guesses (i.e., there is only one password 
matching target password's key-set and length), and it reaches 100\% accuracy with Top-50. For Touch Typists, the accuracy is 
60\% for Top-50, and reaches 90\% with Top-250, which is represents the space reduction of 99,86\% over the original dictionary.

\subsubsection{Using Inter-keystroke Timings}
Although we initially focused on including timings within our password ranking algorithm, this is not straightforward 
due to missing key-pairs in the dataset. We tried to extrapolate the timings for missing key-pairs by creating bins 
for keyboard Euclidean distances of key-pairs and assigning the mean of key-pairs that are in the same bin and 
with known inter-keystroke timings. This approach unfortunately did not work as well as expected, resulting in 
increase in password search space, compared to only using keyboard acoustics.

Ideally, after obtaining timing information on key-pairs that occur in guess passwords, the distribution of inter-keystroke 
timings for each key-pair can be used to compute a probability by modeling each key-pair timing as a normal distribution 
using maximum likelihood estimation. (A similar approach was used in~\cite{song2001timing}.) In future work, we plan to 
incorporate timing information into the \textit{guided search} mechanism.

\subsection{Mitigation Strategies}
There are several simple strategies to mitigate or reduce the threat of \attacka, without modifying any existing hardware. 
The most intuitive solution is to introduce \emph{Chaff typing} right after a password is entered. This can be as simple 
as asking the users to swipe their hands along the keyboard after password entry, or requiring them to introduce noise 
by typing arbitrary ``chaff''. This would serve to obscure the password by introducing useless thermal residues, and thus
make the password key-presses much more difficult to retrieve. 
Another way is to avoid keyboard entry altogether and use the mouse to select (click on) password characters displayed
on the on-screen keyboard. A variation is to have drop-down menu for each position of the password and the user selects 
each character individually. A more burdensome alternative is to use the keyboard arrow keys to adjust a random 
character  string (displayed on the screen) to the actual password. All such methods are well-known and are quite viable. 
However, they are more vulnerable to \surfa, due to the ease of watching a victim's larger, visible screen instead of their smaller, 
partially occluded keyboard.
Finally, a user who is willing to go to extreme lengths to avoid leaving thermal residues could wear insulating gloves 
or rubber thimblettes over their fingers during password entry. This would greatly reduce thermal residues, 
and make \attack ineffective, since thermal conductivity of the insulating material would be much less than that of human skin.
\changed{
Although the effectiveness of such a method depends on the
insulation quality of the material used.
For low insulation materials (e.g., light gloves), the heat transfer could still be significant enough to observe the thermal residues on the image.
	
Note that the visibility of thermal residues depends on how different the temperatures of the residues are from the environment.
Therefore, thermal residues of a person with colder hands still would be visible if the temperature of their fingers are significantly less than that of the keyboard and there is enough contact for the heat to transfer.
In fact, we observed this phenomenon during one of our tests.
The thermal residues left on the keyboard after washing hands were still visible since they significantly differed from the temperature of the keys.
}

If hardware changes are possible, other mitigation techniques might be appropriate. 
For example, a touch-screen would allow password entry without the use of a keyboard. However, this  would be more
(than keyboard entry) vulnerable to \surfa. Also, the use of touch-screens opens the door for attacks that exploit 
smudge patterns left behind by fingers~\cite{aviv2010smudge}.
Alternatively, common plastic keyboards could be replaced with metallic ones. Metals have much higher thermal conductivity 
than plastics. Thus, any localized thermal residues very quickly dissipate throughout the keyboard. A similar strategy was 
adopted to protect ATMs from thermal attacks~\cite{mowery2011heat}.
\changed{Another approach is to place a heat source/plate that regulates an even temperature for all keys throughout the use of the keyboard. Similarly, a random heat pattern could be equally effective. In fact, it has been shown that the hardware that heats up during use on mobile phones could render lock patterns irrecoverable~\cite{andriotis2013pilot} -- laptops might be immune to thermal imaging attacks due to the same reason.
}

To prevent information leakage via 
the acoustic side-channel, more silent keyboards can be used. This reduces the 
chance that 
key presses can be recorded and 
used for key detection later. Random assignments of characters to keyboard 
keys is a possibility (since no key-press sound would 
reveal the actual assignment), however this would also likely increase password 
entry 
time, thus being less user-friendly. Even in this case, 
timings can still reveal information (e.g., frequency analysis by digraphs).
\section{Related Work}
\label{sec:rw}
Human-factors based attacks have been extensively studied over the past decade. 
Interesting side-channels have been discovered~\cite{aviv2010smudge, song2001timing, weinberg2011still} and there has been a wealth of work on strategies to commit and mitigate \surfa~\cite{brudy2014anyone, kumar2007reducing, yamamoto2009shoulder}.
\changed{Even observing hand movements from an opposite perspective (i.e., behind the device screen) and from afar has been shown to be a viable side-channel~\cite{shukla2019stealing} against password entry.
} 

Thermal side-channels have been shown to be an avenue for obtaining secrets (e.g., key-codes, PINs) with the work of Zalewski~\cite{SafeCracking}. Mowery et al.~\cite{mowery2011heat} investigated the influence of material composition (metal vs. plastic) and camera distance (14 vs. 28 inches) on PIN recovery, using a US\$$17,950$ thermal camera, 
on commercial PoS-style PIN pads.~\cite{sidhustudy} explored the effectiveness of a low-cost thermal camera ($\approx US\$330$, attachable to a 
smartphone) to recover 4-digit PINs entered into rubber keypads. Lastly,~\cite{wodo2016thermal} discussed the viability of thermal imaging attacks on various PIN-entry devices including a keyboard, digital door lock, cash machine and payment terminal. Analysis showed that the attack was a credible threat. The attack on keyboards was to recover a 4-digit PIN entry and did not consider passwords.

Androitis et al.~\cite{andriotis2013pilot} investigated using a thermal camera to infer screen-lock patterns of smartphones. Similarly,~\cite{abdelrahman2017stay} conducted more extensive experiments to assess efficacy of thermal imaging attacks against screen-lock patterns. It was shown that PINs were 
vulnerable to such an approach, while swipe-patterns were not.

Additionally, there is a great deal of work showing that keyboard acoustic emanations leak information about pressed keys~\cite{asonov2004keyboard, zhuang2009keyboard, berger2006dictionary, zhu2014context}. \changed{In~\cite{10.1145/2789168.2790122}, the authors recover keystrokes using a single phone placed behind the keyboard, by clustering time-difference of arrival as well as acoustic features.} The effects of typing style (Hunt-And-Peck vs Touch) on the creation of acoustic profiles also have been explored~\cite{halevi2012closer}. Finally, as was recently shown, acoustic emanation attacks can be even mounted remotely~\cite{compagno2017don}.
\section{Conclusions \& Future Work}
\label{sec:conc}
As formerly niche sensing devices become less and less expensive, new side-channel attacks move from ``Mission: Impossible'' 
towards reality. This strongly motivates exploration of novel human-factors attacks, such as those based on \attack. 
Work described in this paper sheds some light on understanding the thermodynamic relationship between human fingers 
and external computer keyboards. In particular, it exposes the vulnerability of standard password-based systems to 
adversarial collection of thermal emanations.

Based on the study results, we believe that \attacka represent a new credible threat for password-based systems, 
and that human-induced  thermal side-channels deserve further study. This is especially true considering 
constantly decreasing costs and increasing availability of high-quality thermal imagers.  It is realistic to expect that 
-- in several years' time -- thermal imagers that can be attached to smartphones, e.g., FLIR One (Figure~\ref{fig:TG165}), will offer the quality 
equivalent to SC620 that was used in our study. This would allow surreptitious collection of thermal images without 
bulky, unusual or suspicious-looking equipment. Cameras in the price range of our SC620 would offer the image quality 
of A6700sc, with time-windows for collecting thermal residues that last for several minutes. 

Although thermal residue offers a prominent side-channel to insider attackers for password recovery, it is not without its limitations. To circumvent such limitations, insider threats will always look for ways to increase their success of recovering passwords without exposing their actions. To this end, combination of multiple side-channels offers an avenue. Therefore, we anticipate 
the following future work directions: 
%
\begin{compactitem}
\item Given marked differences in collectible data between Touch and Hunt-and-Peck typists, one interesting next step is to 
further refine our attack to handle expert typists who introduce natural chaff through resting their hands on the 
keyboard home-row. Correct disambiguation of a home-row key being a part of the password rather than thermal noise,
would be very helpful in limiting the password search space. 
\item Another future direction is a longitudinal study to model multiple instances of \attacka, i.e., where the adversary, over time,
has several chances to obtain thermal imaging data against the same victim. Our study only measured thermal residues 
from each subject once, per password per keyboard. We hypothesize that a more persistent adversary would
be more successful and would be more likely to recover the entire password after multiple \attack instances. 
However, substantial further experiments are needed to substantiate this claim. 
\item It would also be useful to investigate lowering the bar for adversarial sophistication. 
Figure~\ref{fig:TG165} shows an image of password {\sf ``football''} entered by a Hunt-and-Peck typist, 
$10$ seconds after entry, as captured by the inexpensive FLIR ONE -- a low-tier thermal camera attachable to a mobile phone. This image 
suggests that, in the long run, even a less capable (in terms of equipment) adversary may pose a credible threat.
%

\item We plan to improve upon \acutherm\ by exploring additional fusion strategies of multiple side-channels. For example, side-channels 
can provide different levels of information and, depending on their accuracy, a weighted sum can be used to achieve better results. 
The weights can be learned by a machine learning algorithm, based on previous data.

\item Finally, we plan to investigate how to incorporating timings into our current results with a more complete 
dataset that includes all key-pairs.

\end{compactitem}

\bibliographystyle{unsrt}  
\bibliography{main}

\end{document}